\documentclass[twocolumn]{aastex631}
\usepackage{graphicx}

\usepackage{ulem}
\usepackage[T1]{fontenc}

\newcommand{\sub }[1]{_{\rm{#1}}}

\newcommand{\rp}{r\sub{p}}
\newcommand{\hp}{h\sub{p}}
\newcommand{\cs}{c\sub{s}}
\newcommand{\MJ}{M\sub{J}}
\newcommand{\Mp}{M\sub{p}}
\newcommand{\Mstar}{M\sub{\ast}}
\newcommand{\Msun}{M\sub{\odot}}
\newcommand{\RH}{R\sub{H}}
\newcommand{\Mdot}{\dot{M}\sub{p}}

\shorttitle{Eccentric gap by a super-Jupiter mass planet}
\shortauthors{Tanaka et al.}
\graphicspath{{./}{figures/}}

\begin{document}

\title{Eccentric gap induced by a super-Jupiter mass planet}

\correspondingauthor{Yuki A. Tanaka}
\email{yuki.tanaka@astr.tohoku.ac.jp}

\author[0000-0002-0141-5131]{Yuki A. Tanaka}
\affiliation{Astronomical Institute, Tohoku University, Sendai, Miyagi 980-8578, Japan}

\author[0000-0001-7235-2417]{Kazuhiro D. Kanagawa}
\affiliation{College of Science, Ibaraki University, 2-1-1 Bunkyo, Mito, Ibaraki 310-8512, Japan}

\author[0000-0001-9659-658X]{Hidekazu Tanaka}
\affiliation{Astronomical Institute, Tohoku University, Sendai, Miyagi 980-8578, Japan}

\author[0000-0002-5964-1975]{Takayuki Tanigawa}
\affiliation{National Institute of Technology, Ichinoseki College, Ichinoseki, Iwate 021-8511, Japan}




\begin{abstract}

A giant planet embedded in a protoplanetary disk opens a gap by tidal interaction, and properties of the gap strongly depend on the planetary mass and disk parameters.
Many numerical simulations of this process have been conducted, but detailed simulations and analysis of gap formation by a super-Jupiter mass planet have not been thoroughly conducted.
We performed two-dimensional numerical hydrodynamic simulations of the gap formation process by a super-Jupiter mass planet and examined the eccentricity of the gap.
When the planet is massive, the radial motion of gas is excited, causing the eccentricity of the gap's outer edge to increase.
Our simulations showed that the critical planetary mass for the eccentric gap was $\sim3~\MJ$ in a disk with $\alpha=4.0\times10^{-3}$ and $h/r=0.05$, a finding that was consistent with that reported in a previous work.
The critical planetary mass for the eccentric gap depends on the viscosity and the disk scale height.
We found that the critical mass could be described by considering a dimensionless parameter related to the gap depth.
The onset of gap eccentricity enhanced the surface density inside the gap, shallowing the gap more than the empirical relation derived in previous studies for a planet heavier than the critical mass.
Therefore, our results suggest that the mass accretion rate, which strongly depends on the gas surface density in the gap is also enhanced for super-Jupiter mass planets.
These results may substantially impact the formation and evolution processes of super-Jupiter mass planets and population synthesis calculations.

\end{abstract}




\section{Introduction}\label{introduction}


A planet embedded in a protoplanetary disk gravitationally establishes interaction with the surrounding gas.
If the planet is heavier than several tens of Earth masses, it leads to the opening of a low-density gap structure along with its orbit \citep[e.g.,][]{lp79,gt80,al94,kle99,cri06}.
Since the opening of the gap exerts considerable impacts on planetary growth and migration \citep{lp79}, detailed investigations of the nature of the gap are important to understand planet formation, particularly for gaseous planets.

Classical gap models assumed that there is almost no gas inside the gap \citep{lp86}, suggesting that planet growth ends when the gap is opened because there will be no mass accretion.
However, later hydrodynamical simulations have shown that gas can flow across the gap, suggesting that the mass accretion onto the gaseous planet continues even after the gap opens \citep[e.g.,][]{kle99,lub99,ms01,dan03,mac10,zhu11}.
Recently, several numerical studies have reported the examination of structures such as depth and width, and an empirical model of the gap induced by an embedded planet has been provided \citep[e.g.,][]{dm13,fun14,kan15a,kan16,kan17,dc15,fc16}.
Based on this empirical model of the surface density of the gap, an analytic formula for estimating the mass accretion rate onto the planet was derived by \citet{tt16}, and their finding was consistent with hydrodynamical simulations \citep{dan03,mac10}.
Such models describing the gap structures and the accretion rate onto the gaseous planet were used in population synthesis calculations performed for exoplanets \citep[e.g.,][]{ida18,tan20}.

The disk–planet interaction is also important for interpreting observations of protoplanetary disks.
Recent observations, mainly achieved by using the Atacama Large Millimeter/submillimeter Array (ALMA) telescope, have discover well-defined structures of multiple rings and gaps in a substantial number of protoplanetary disks \citep[e.g.,][]{alm15,and16,ise16,cie17,loo17,fed18}.
Gaseous planets embedded in the disk are highlighted as one of the promising explanations for these rings and gaps, and many numerical studies on the disk–planet interaction have been conducted \citep[e.g.,][]{dip15,don15,kan15a,kan15b,kan16,kan17,pk15,jin16}.
Several possible formation paths of the planets at $\sim100$ au in an early phase have been investigated \citep{dd18,joh19,tan19}.

When the planet embedded in the protoplanetary disk is adequately massive, the gap induced by the planet presents with significant eccentricity \citep{kd06}.
In their simulations, \citet{kd06} have suggested the following: 1) if a planet is less massive than $\sim3~\MJ$, it will induce a circular gap; 2) if a planet is more massive than $\sim3~\MJ$, the inner edge of the gap will not exhibit evident eccentricity; however, the outer edge will present a significant eccentricity of approximately 0.2; and 3) mass accretion rate onto the planet is considerably enhanced when the planet mass is heavier than $5~\MJ$ because of the effect of the gap's eccentricity.
Several other studies also investigated instabilities near the gap region and the onset of eccentricity \citep[e.g.,][]{kol03,go06,fun14}.

However, previous studies present a discrepancy in the mass accretion rate onto the super-Jupiter mass planets.
As noted above, \citet{kd06} predicted that the mass accretion rate was enhanced in case of the super-Jupiter mass planets.
On the other hand, based on numerical simulations, \citet{bod13} showed that the mass accretion rate would be much lower than the empirical relations for a planet heavier than $3~\MJ$.
This discrepancy suggests that our understanding of gap formation by a super-Jupiter mass planet and the mass accretion onto the planet remain elusive and inadequate.

Since the previous empirical models proposed for the gap structure and the mass accretion rate onto the planet are based on the assumption that the gap is almost circular, these models should be modified if the planet is massive enough to induce an eccentric gap.
This may also affect the migration of the planet because the torque exerted on the planet depends on the surface density in the gap \citep{kan18}.
The modification of the model of the gap structures induced by a super-Jupiter mass planet may impact the population synthesis models of exoplanets, including empirical models of the gap structures and the mass accretion rate \citep[e.g.,][]{tan20}.
Therefore, a comprehensive study of the disk–planet interaction using hydrodynamical simulations for the super-Jupiter mass regime is necessary to formulate and establish the planet formation theory.

The purpose of this study is to investigate gap formation in a protoplanetary disk by a super-Jupiter mass planet in wide parameter ranges and to reveal gap structure parameter dependencies, such as the depth and width of the gap and the disk eccentricity, using two-dimensional hydrodynamical simulations.
Our findings show that if the planet is massive enough, the radial motion of the disk gas is excited, and the outer edge of the gap gains eccentricity.
The critical mass of the planet to induce an eccentric gap depends on the viscosity and the scale height of the disk.
Our findings also highlight the condition necessary for the onset of the eccentricity of the gap and present a comparison of the resultant gap structures with the empirical relation derived in the previous studies to elucidate the mechanism by which the eccentricity of the gap affects conventional gap models.

This paper is organized as follows.
In Section \ref{method}, we have briefly described the setup of our numerical simulations.
In Section \ref{result}, we have presented main findings based on the simulations such as the distributions of the surface density, disk eccentricity, gap depth, and their time and parameter dependences.
We also redefine the eccentricity of the disk to evaluate the transition from a circular gap to an eccentric gap more accurately.
Then, we have provided evidence that the dependence of the gap depth and the transition to the eccentric gap can be well described by considering a dimensionless parameter that depends on the mass of the planet, the viscosity, and the scale height of the disk.
Additionally, we have examined the relation between the gap width and the eccentricity of the gap.
Section \ref{discussion} presents discussions on the relation of the eccentricity and the surface density at resonances, the mass accretion rate onto the super-Jupiter mass planet, and implications for observations.
We have also briefly mentioned the torque exerted on the planet and its parameter dependence in the section.
Finally, we have summarized the findings of our work in Section \ref{summary}.


\section{Numerical Method}\label{method}


We numerically investigated the properties of the gap created by a planet embedded in a protoplanetary disk and its parameter dependencies using the hydrodynamic code FARGO \citep{mas00}.
This code is used to solve the equations of continuity and motion of disk gas and is widely used to calculate disk--planet interaction \citep[e.g.,][]{cm07,bar11,zhu11}.
We considered a two-dimensional cylindrical coordinate $(r,\,\phi)$, and inserted a central star at $r=0$, and the orbital distance of a planet was fixed to $r=\rp$ with a circular orbit.
The computational domain ranged from $r/\rp=0.4$ to 4.0 in all calculations.
The resolution was $512\times1536$ for radial and azimuthal zones, and we adopted logarithmic spacing for the radial direction and equally spacing for the azimuthal direction.
We considered a constant kinematic viscosity $\nu$, wherein $\nu=\alpha\cs\hp$, with $\cs$ and $\hp$ representing sound speed and a disk scale height at $r=\rp$, respectively, and $\alpha$ denotes the conventional $\alpha$-prescription of \citet{ss73b}.
We also assumed that the aspect ratio of the disk $h/r$ was constant throughout the disk.
The disk was assumed to be geometrically thin, non-self-gravitating, and locally isothermal.
A smoothing length of the gravitational potential of the planet was set to $0.6\hp$.

To investigate the gap formation by a super-Jupiter mass planet in the protoplanetary disk, we considered variations in the the mass ratio of a planet and a central star from $10^{-3}$ to $10^{-2}$, corresponding to $\Mp=1$ -- $10~\MJ$, if the mass of the central star was assumed to be $\Mstar=1~\Msun$.
For disk parameters, we adopted $\alpha=4.0\times 10^{-3}$ and $h/r=0.05$ as a fiducial case, and $\alpha=10^{-3}$, $10^{-2}$ and $h/r=0.07$, $0.1$ were considered to evaluate the dependence on the viscosity and the aspect ratio of the disk.
The viscosity and the aspect ratio were set and fixed to constant values throughout the disk.

An initial profile of the surface density of the gas $\Sigma$ was set to
\begin{eqnarray}
\Sigma\!\left(r\right)=\Sigma\sub{0}\left(\frac{r}{\rp}\right)^{-1/2},
\end{eqnarray}
where $\Sigma\sub{0}$ represents the surface density at the orbit of the planet.
The initial radial velocity of the gas is expressed as
\begin{eqnarray}
v_{r}=-\frac{3\nu}{2r},
\label{eq:vr}
\end{eqnarray}
and the initial azimuthal velocity is expressed as
\begin{eqnarray}
v_{\phi}=v\sub{K}\sqrt{1-\eta},
\label{eq:vphi}
\end{eqnarray}
where $v\sub{K}=\sqrt{GM\sub{*}/r}$ denotes the Keplerian velocity, $G$ represents the gravitational constant, and $\eta$ is expressed as
\begin{eqnarray}
\eta=\frac{1}{2}\left(\frac{h}{r}\right)^{2}\frac{d\ln P}{d\ln r},
\end{eqnarray}
where $P$ represents the gas pressure.
We used a mass-taper function for the mass of the planet, where the mass smoothly increased from zero to the final value using the mass-taper function defined by $\sin^{2}\left[\pi t/\left(20 P\sub{orbit}\right)\right]$, where $P\sub{orbit}$ denotes the orbital period of the planet.
To focus on properties of the gap structure and their dependence on the parameters of the planet and the disk, we neglected the mass accretion onto the planet, such that the mass of the planet was fixed.
The orbit of the planet was fixed, and the planetary migration was also neglected for the same reason.

At the inner and outer boundaries, we adopted a viscous boundary condition where the radial and azimuthal velocities of the gas were described by Equations (\ref{eq:vr}) and (\ref{eq:vphi}), which was also used in the studies reported by \citet{kan16,kan17}.
Additionally, we used wave-killing zones near the inner and outer boundaries ($0.4<r/\rp<0.5$, and $3.5<r/\rp<4.0$, respectively) to avoid artificial wave reflection at the boundaries \citep{dev06}.
An advantage of the existence of a viscous boundary condition was that it can help avoid unphysical outflow of the material at the boundaries; the consideration of such an aspect was particularly important for calculations of the disk with a massive gaseous planet.
One of the possible boundary conditions, an open boundary, allowed gas to flow outside the computational domain if $v_{r}<0$ at the inner boundary.
This boundary condition is valid for a less massive gaseous planet because density waves excited by the planet possess a smaller amplitude.
In this condition, however, unphysical loss of the gas from the computational domain can occur when density waves excited by the gravity of the planet reach the inner boundary if the planetary mass is larger.
Although it can be generally suppressed using the wave-killing zones to dampen the density waves near the boundaries, its effect will not be neglected for a super-Jupiter mass planet because as the amplitude of the density waves continues to increase, wave-killing zone insufficiency increases, causing unrealistic disk accretion to the inside of the inner boundary and depletion of the gas in the inner region of the disk.

\begin{figure}[htbp]
  \centering
  \includegraphics[]{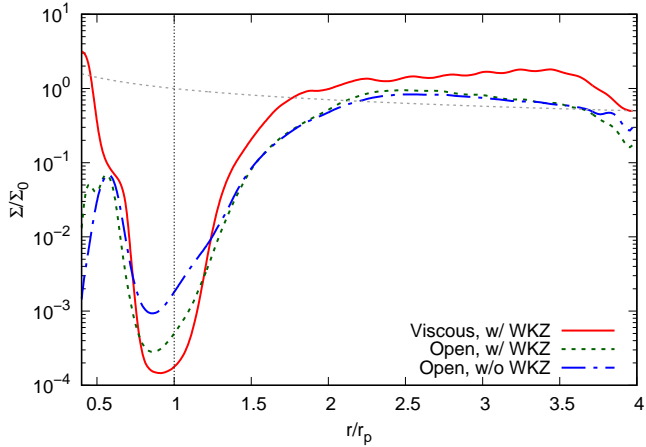}
  \caption{
  Comparison of the radial surface density profiles around the gaps with different boundary conditions and wave-killing zones.
  Parameters for the disk and the planet were fixed in these calculations, as $\Mp=10\MJ$, $\alpha=4\times10^{-3}$, and $h/r=0.05$.
  The surface densities were time- and azimuthally averaged without considering the vicinity of the planet and were normalized by considering the initial value at $r=\rp$.
  The time average begins from $t=1500$ to 2000 orbits.
  The solid red line represents the viscous boundary with wave-killing zones scenario.
  The green dotted lines represents the open boundary with wave-killing zones scenario.
  The blue dash-dotted line represents the open boundary without the wave-killing zones scenario.
  The gray dotted line corresponds to the initial surface density, and the vertical black dotted line shows the location of the planet.
  }
  \label{fig:boundary}
\end{figure}
Figure \ref{fig:boundary} shows the effects of the boundary conditions and the wave-killing zones on the surface density profile of the disk when the mass of the planet is $10~\MJ$.
For open boundary scenarios without wave-killing zones (blue dash-dotted line), the surface density in the vicinity of the inner boundary decreased several orders of magnitude compared to the viscous boundary scenarios with wave-killing zones (solid red line).
When we included wave-killing zones, depletion at the inner boundary was relatively suppressed (green dotted line) because the density waves excited by the planet were dampened in the wave-killing zones.
The artificial outflow was decreased, but it largely deviated from the viscous boundary with wave-killing zones scenario.
Figure \ref{fig:boundary} also shows that the surface densities around the inner boundary and outside the gaps decreased compared to the viscous boundary scenario.
These effects were weaker when the mass of the planet was relatively small because the amplitude of the density waves excited by the planet was small, and material loss at the inner boundary was suppressed.
For heavier planets, however, the effects cannot be neglected, as shown in Figure \ref{fig:boundary}.
The surface densities near the planet's orbit differ by an order of magnitude depending on the boundary conditions and the wave-killing zone.
In the case of the open boundary without the wave-killing zone (blue dash-dotted line), eccentric motion of the gas inside the gap was enhanced compared to the other cases, thus the surface density near the planet's orbit became larger.
Since the surface density around the gap strongly affects the accretion rate onto the planet and the migration rate, the consideration of the boundary condition and the wave-killing zones is particularly important for the cases of massive planets like super-Jupiter mass planets.

Since the origin of the computational domain was fixed to the position of the central star, we also included the indirect term of the gravitational potential to treat the gravitational interaction between the central star and the planet properly.
The inclusion of the indirect term is also particularly important for massive planets because the reflex motion of the central star is substantial when the planet is large, and it can affect the properties of the gap structure created by the planet.


\section{Result}\label{result}



\subsection{Fiducial cases}\label{result_fiducial}



\subsubsection{Transition to eccentric gaps in fiducial disks}\label{result_fiducial_transition}


\begin{figure*}[htbp]
  \centering
  \includegraphics[width=5.5cm]{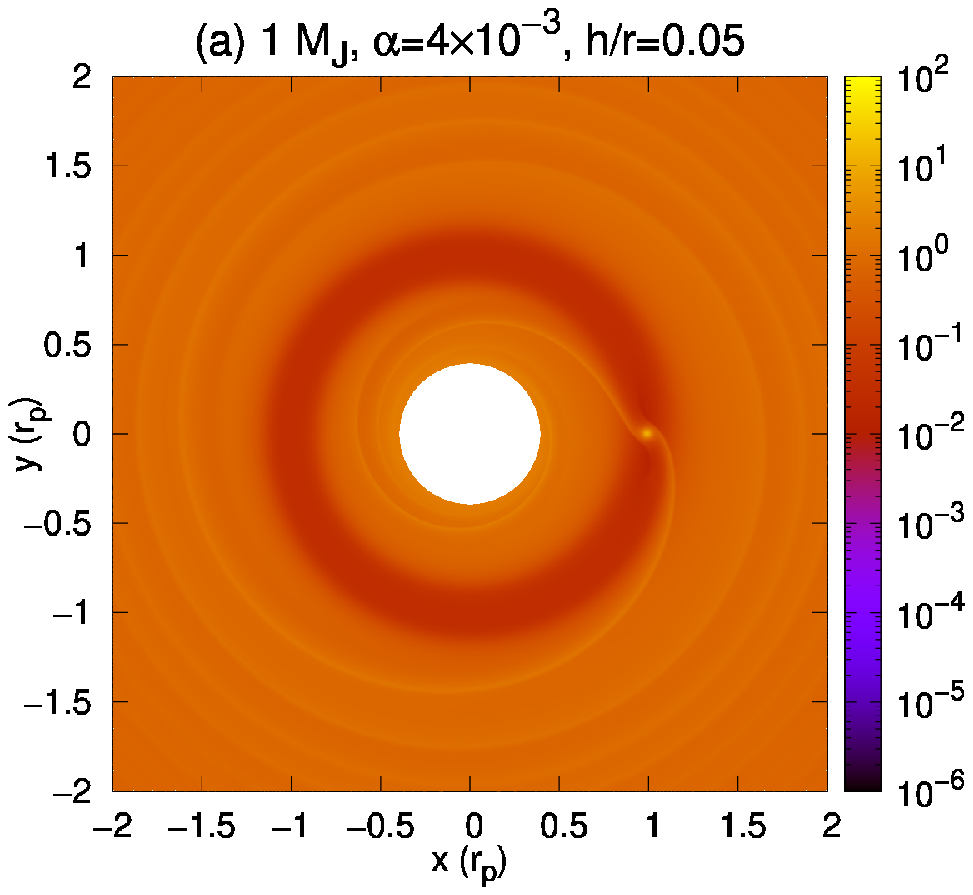}
  \includegraphics[width=5.5cm]{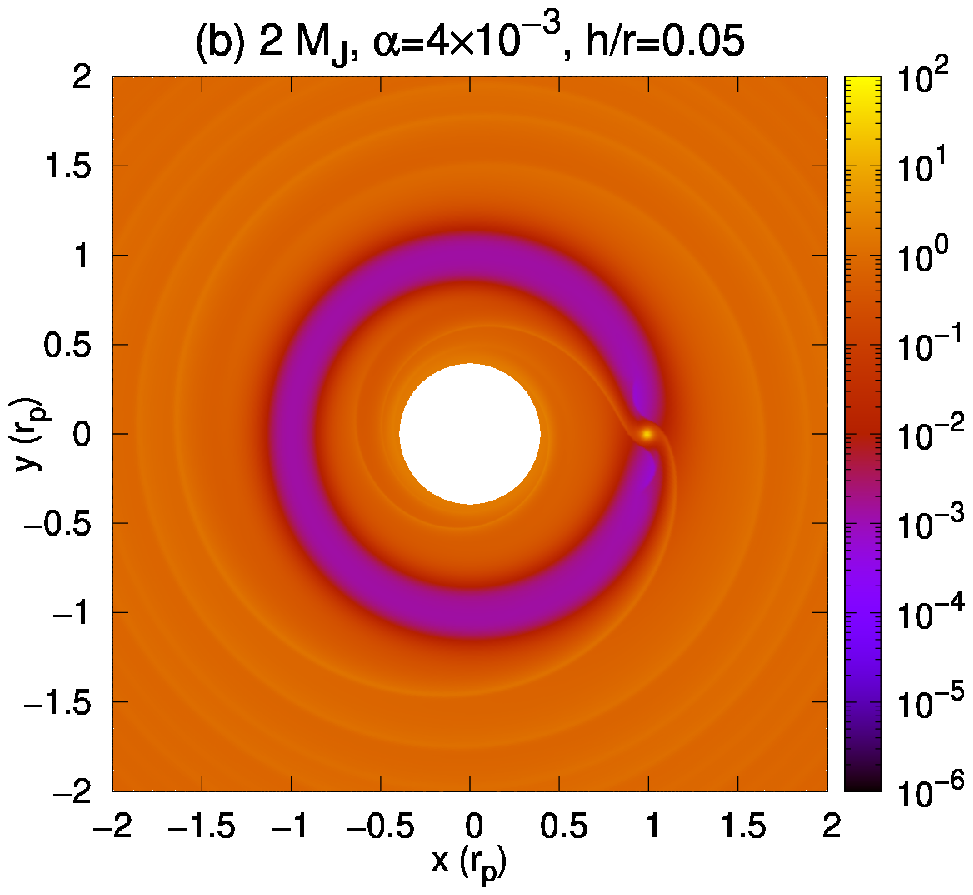}
  \includegraphics[width=5.5cm]{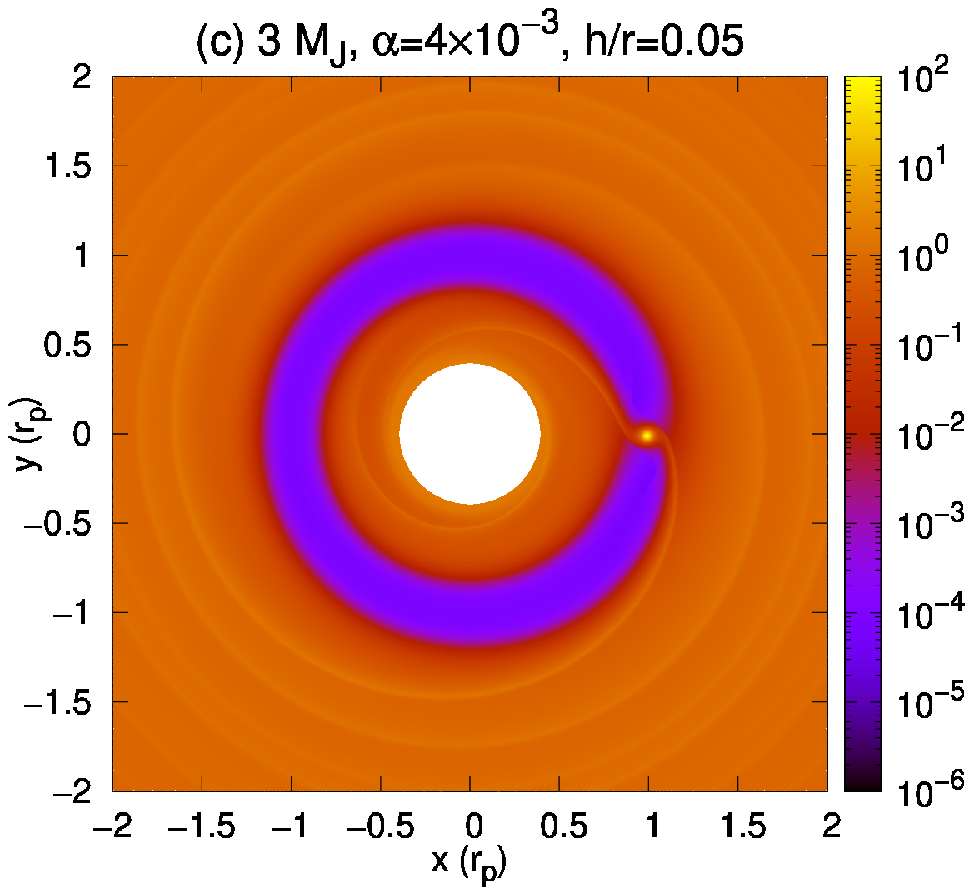}
  \includegraphics[width=5.5cm]{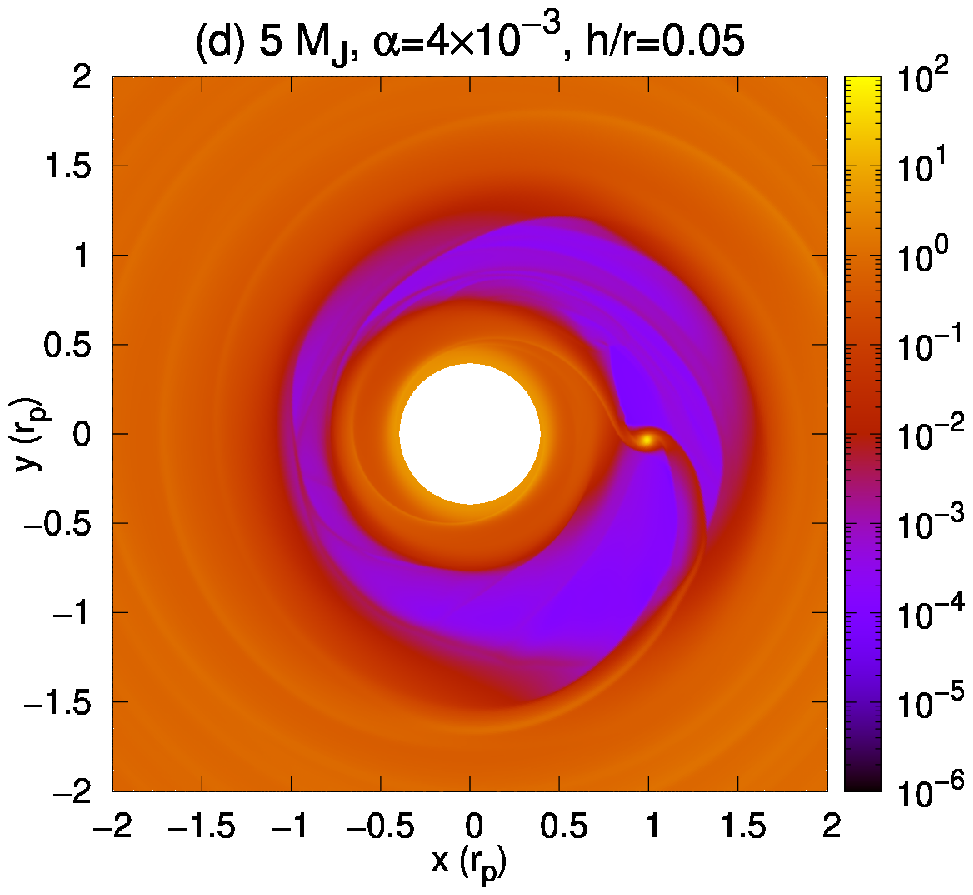}
  \includegraphics[width=5.5cm]{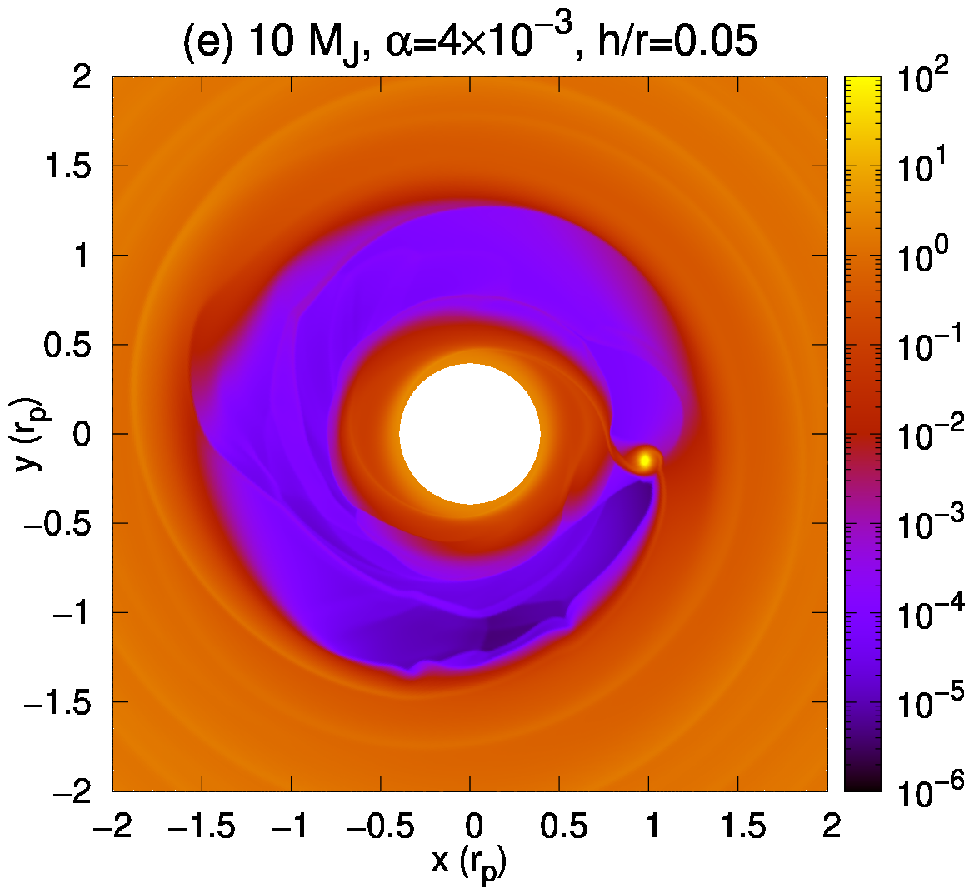}
  \caption{
  The surface density distribution at $t=2000$ orbits.
  The disk parameters were set to $\alpha=4\times 10^{-3}$ and $h/r=0.05$.
  Planetary masses are (a) 1~$\MJ$, (b) 2~$\MJ$, (c) 3~$\MJ$, (d) 5~$\MJ$, and (e) 10~$\MJ$.
  The surface density was normalized by considering the initial (unperturbed) value at $r=\rp$ and has been shown in logarithmic color scale.
  }
  \label{fig:sigmamap1}
\end{figure*}
First, we have presented the results with the fiducial disk parameters ($\alpha=4\times10^{-3}$ and $h/r=0.05$).
Figure \ref{fig:sigmamap1} shows the surface density distributions at $t=2000$ planetary orbits with $\Mp=1$ -- $10~\MJ$.
For 1 -- $3~\MJ$ scenarios, the gap shape created by the planet was almost axisymmetric (panels (a) to (c)).
The surface density inside the gap seemed to be almost constant and showed no time variations.
It has been shown that the surface density of the gap region decreased as the mass of the planet increased.
For heavier planets, however, the appearance of the gap changed significantly (panels (d) and (e)).
The outer edge of the gap showed eccentricity, and the surface density of the gap region displayed considerable spatial and time variabilities.
Note that the gap whose outer edge exhibits significant eccentricity has been referred to as an eccentric gap.
Although small time variations can be observed, an inner edge of the gap does not show significant eccentricity even if the outer edge shows eccentricity.
This may be caused by asymmetric nature between the inner and outer edges.
Because the deep gap induced by a heavier planet has a wider gap width, the asymmetry between the inner and outer edge of the gap will be likely to appear.
We investigate this point in terms of the condition of Rayleigh instability in Section \ref{resonance}.
In addition to this, it is possible that the proximity of the inner boundary to the gap creates the difference between the inner and outer edges of the gap.
In order to verify this, calculations with a wider inner boundary is needed, which will be a future task.

\begin{figure*}[]
  \centering
  \includegraphics[width=5cm]{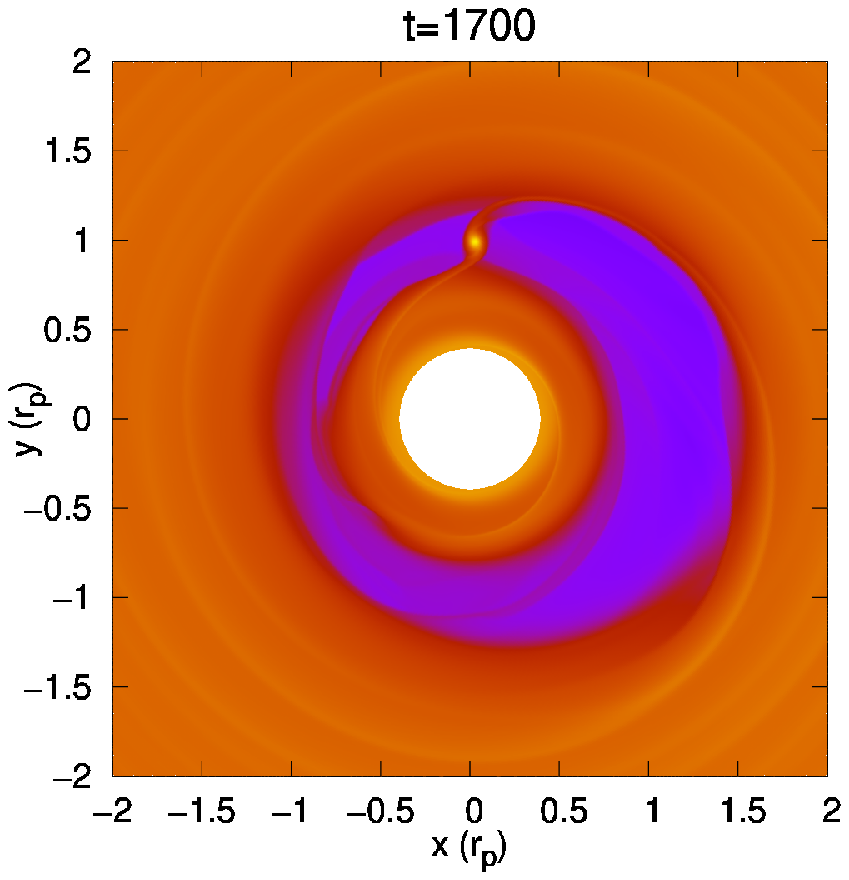}
  \includegraphics[width=5cm]{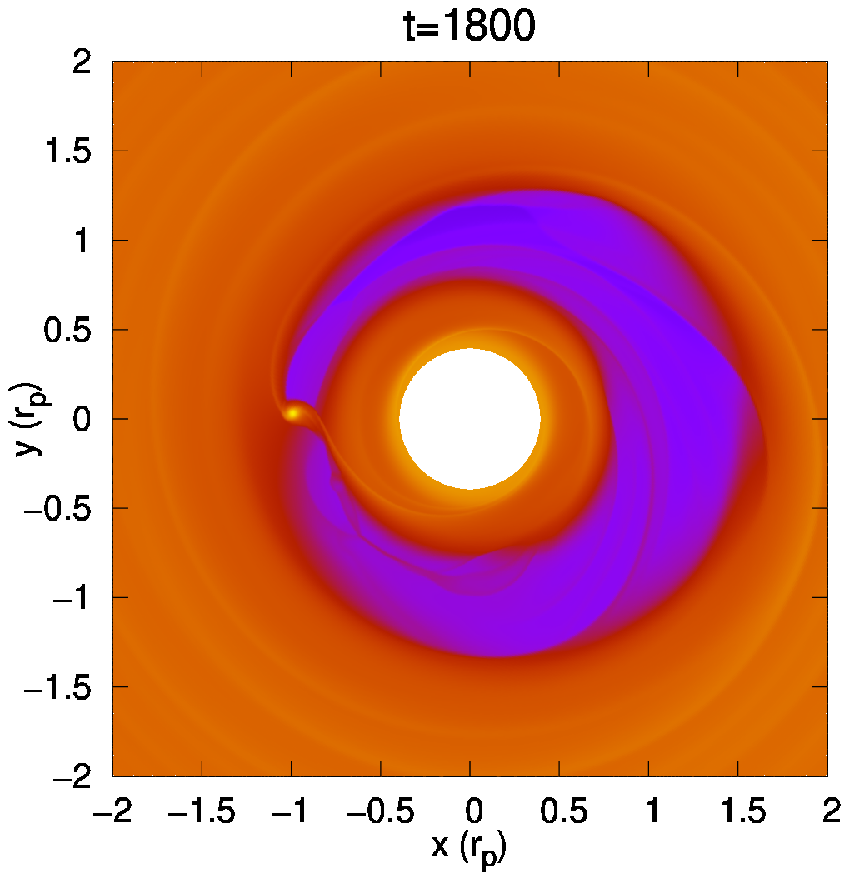}
  \includegraphics[width=5cm]{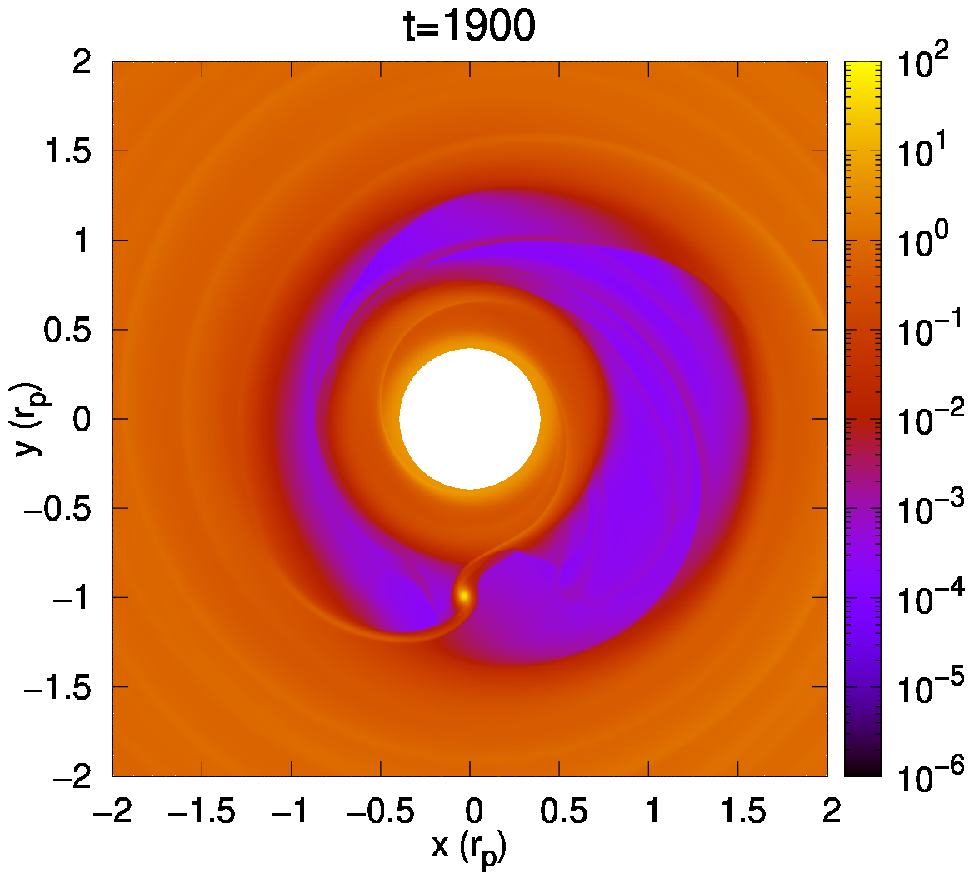}
  \caption{
  The snapshots depicting the surface density distribution for the case of $5~\MJ$ for $t<2000$ orbits.
  The axes and color scale are the same as Figure \ref{fig:sigmamap1}.
  }
  \label{fig:sigmamap_time}
\end{figure*}
In cases exhibiting eccentric gap scenarios, relative configuration between the eccentric outer edge and the planet changes with time.
Figure \ref{fig:sigmamap_time} shows an example of time evolution for the relative location when $\Mp=5~\MJ$.
At $t=1700$ (left panel), the planet was located in the direction of the minor axis of the eccentric gap.
At $t=1800$ (middle panel), the planet was located near the pericenter.
At $t=1900$ (right panel), the planet moved to the minor axis again.
Finally, at $t=2000$ (Figure \ref{fig:sigmamap1} (d)), the planet moved near the apocenter of the eccentric gap.
These suggest that the surface density in the gap near the planet will vary with time significantly depending on the relative location, and it will affect the mass accretion rate onto the planet for the heavier planets.
We have discussed this aspect in Section \ref{mdot}.

\begin{figure}[]
  \centering
  \includegraphics[]{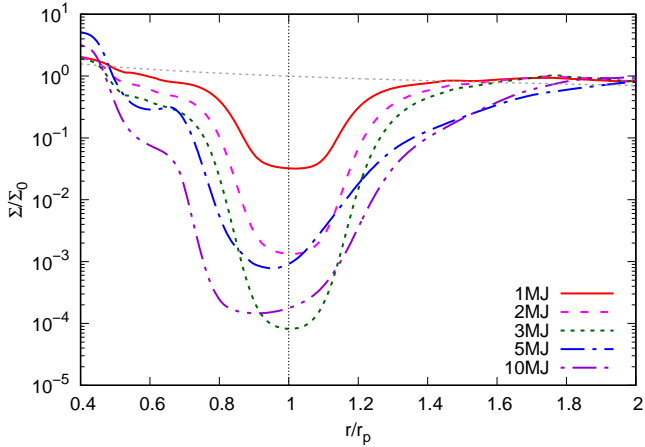}
  \caption{
  The radial distribution of the azimuthally and time-averaged surface densities around the planets with various masses.
  Solid red, dashed magenta, dotted green, dash-dotted blue, dashed double-dotted violet lines correspond to 1, 2, 3, 5, and 10~$\MJ$, respectively.
  The dotted grey line shows the initial surface density, and the vertical black line shows the location of the planet.
  The surface densities were normalized by considering the initial (unperturbed) value $\Sigma_{0}$ at $r=\rp$.
  }
  \label{fig:sigmaave1}
\end{figure}
Since mass accretion onto the planet in the simulations was neglected, an envelope-like density distribution was observed due to the planet's gravity, as shown in Figures \ref{fig:sigmamap1} and \ref{fig:sigmamap_time}.
We excluded the planet's vicinity when the surface density was azimuthally averaged to avoid this artificial density structure.
The averaged time ranged from $t=1500$ to 2000 because the gap structure could be assumed to reach a steady state condition (or quasi-steady-state for heavier planets) after 1500 orbits.

Figure \ref{fig:sigmaave1} shows the different surface density radial distributionos around the gap between the 1 -- $3~\MJ$ cases and 5, $10~\MJ$ cases.
For the 1 -- $3~\MJ$ cases, the gap structures are almost symmetrical with respect to the vertical dotted line that corresponds to planetary orbit location.
For the 5 and $10~\MJ$ cases, however, the averaged radial distributions are asymmetrical with respect to the planetary orbit, and the deepest point of the gap is located inside the orbit.
This asymmetric distribution reflects the eccentric nature of the gap created by the heavier planets.
It also shows that the effects of the eccentricity are less considerable at the inner edge of the gap; however, considerable effects were observed at the outer edge.
As shown in Figures \ref{fig:sigmamap1} and \ref{fig:sigmamap_time}, the inner edge of the gap does not exhibit significant eccentricity.
Therefore, the slopes of the gap inner edge shown in Figure \ref{fig:sigmaave1} were similar in all cases, but the shallowness of the slopes of the outer edge increased in the heavier planet cases compared to the non-eccentric gaps.
The dependence of the depth of the gap also changed.
For example, the depth of the gap by the $5~\MJ$ planet was shallower than the case of $3~\MJ$, and the depth of the $10~\MJ$ case was comparable to the case of $3~\MJ$.


\subsubsection{Disk eccentricity and its redefinition}\label{result_fiducial_ecc}


\begin{figure}[]
  \centering
  \includegraphics[]{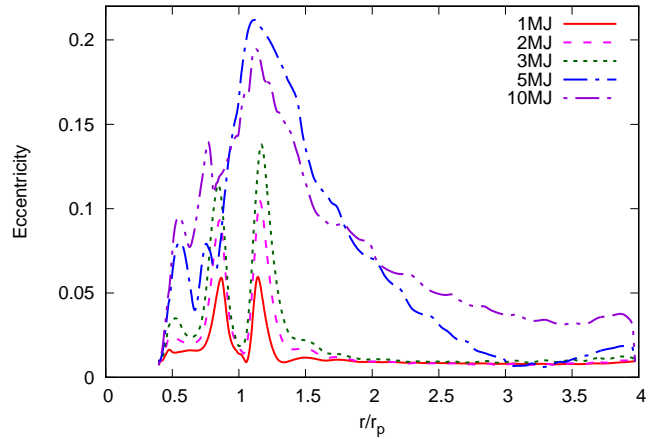}
  \caption{The radial distribution of eccentricities of gas disks for cases of $\alpha=4.0\times10^{-3}$ and $h/r=0.05$ at $t=2000$ orbits.
  Lines represent the same values as in Figure \ref{fig:sigmaave1}.
  }
  \label{fig:ecc}
\end{figure}
For discussion of the gap structure and its parameter dependence, it is important to evaluate the radial profile of the eccentricity of the gas disk.
We performed an eccentricity measurement used by \citet{kd06} for the comparison.
The eccentricity vector of a gas element is expressed as
\begin{eqnarray}
\bf{e}=\left(\frac{\left|\bf{v}\right|^{2}}{\mu}-\frac{1}{\left|\bf{r}\right|}\right)\bf{r}-\frac{\bf{r}\cdot\bf{v}}{\mu}\bf{v},
\end{eqnarray}
where $\bf{r}$ and $\bf{v}$ denote the position and velocity vectors, and $\mu=G\Mstar$ represents the standard gravitational parameter.
The eccentricity is calculated by considering $e=\left|\bf{e}\right|$.
Therefore, in the two-dimensional cylindrical coordinate, the eccentricity can be calculated from the position and velocity and can be expressed as
\begin{eqnarray}
e=\sqrt{\left(\frac{r v_{\phi}^{2}}{\mu}-1\right)^{2}+\left(\frac{r v_{r} v_{\phi}}{\mu}\right)^{2}}.
\label{eq:ecc}
\end{eqnarray}
We calculated the eccentricity of each computational cell using Equation (\ref{eq:ecc}), assuming that the cells were orbiting the central star without being subjected to the pressure gradient force.
Then we average over the eccentricities of every cell in azimuth direction to obtain the radial profile of the disk eccentricity.

Figure \ref{fig:ecc} shows the radial profiles of the disk eccentricity at $t=2000$ orbits.
When the mass of the planet was smaller than $3~\MJ$, the profiles presented with double peaks at the sides of the planet's orbit, and eccentricity was nearly 0.01 in areas away from the planet, indicating that the eccentricity value was dominated by the gap edges for non-eccentric gap scenarios.
On the other hand, the profiles remarkably changed when the mass of the planet is large.
Regarding cases with 5 and $10~\MJ$, the double-peak profile was not observed.
Instead, they exhibited a significant peak around $r/\rp\sim1.2$, with a wide distribution outside the planet's orbit.
The eccentricities at the peak of the distribution were approximately 0.2 for both 5 and $10~\MJ$ cases, implying that the eccentricity of the outer edge of the gap did not strongly depended on the mass of the planet.

The characteristics of the disk's eccentricity profile are shown in Figure \ref{fig:ecc}.
The double-peak distribution for the non-eccentric gap cases and the significant peak outside the orbit with a wide distribution for the cases of the eccentric gap are qualitatively similar to those reported by \citet{kd06}.
Similar radial distributions of the disk eccentricity were also shown by \citet{li21}.
The results for the case of $3~\MJ$ were different between \citet{kd06} and ours; the profile described by \citet{kd06} was similar to the case of the eccentric gap.
This might be caused by differences in numerical setup, for example, the resolution or numerical schemes.
We also considered a case of $\Mp=3~\MJ$ without using the indirect term as a reference and found that the gap's eccentricity increased in contrast to the case considered with the indirect term.
We also noted that our calculation for the $3~\MJ$ case did not transition into the eccentric gap after $t=2000$ orbits.
We extended the calculation up to $t=4000$ orbits and found that the depth of the gap and the eccentricity profile did not change compared to those at $t=2000$ orbits.

In the eccentricity profile of the disk shown in Figure \ref{fig:ecc}, there is a certain eccentricity value at the double peaks at sides of the planetary orbit even in the non-eccentric cases of 1 -- $3~\MJ$.
As shown in Figure \ref{fig:sigmamap1}, however, both the inner and outer edges of the gap do not show eccentricity.
This is attributable to the evaluation of the eccentricity in Equation (\ref{eq:ecc}).
Since the Kepler velocity is written as $v\sub{K}=\sqrt{\mu/r}$, Equation (\ref{eq:ecc}) can be rewritten as follows:
\begin{eqnarray}
e=\sqrt{\left\{\left(\frac{v_{\phi}}{v\sub{K}}\right)^{2}-1\right\}^{2}+\left(\frac{v_{r}}{v\sub{K}}\right)^{2}\left(\frac{v_{\phi}}{v\sub{K}}\right)^{2}}.
\label{eq:ecc2}
\end{eqnarray}
The first term in the square root of Equation (\ref{eq:ecc2}) represents the deviation of a fluid element from the Kepler velocity; if the fluid element moves with the Kepler velocity, the value of this term becomes zero.
However, velocities of fluid elements often deviate from the Kepler velocity because they are subjected to the pressure gradient force as well as the gravity of the central star; hence, the value of this term does not become zero if the gas rotate in a circular orbit.
This deviation is particularly substantial around the gap region because of the immense pressure gradient at the gap edges.
Therefore the double-peak distribution of the eccentricity is observed around the planet's orbit in the non-eccentric gap scenarios shown in Figure \ref{fig:ecc}.

\begin{figure}[]
  \centering
  \includegraphics[]{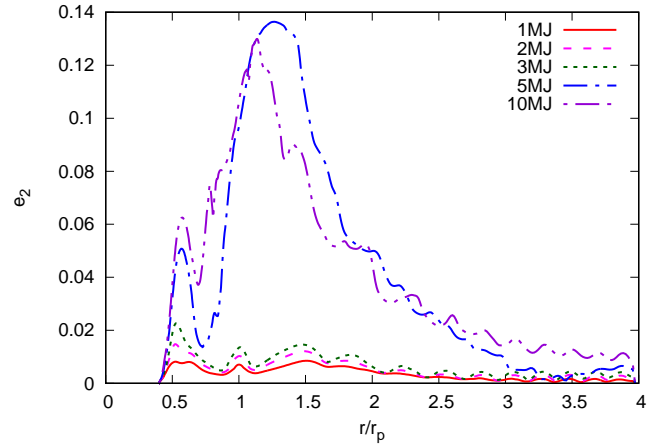}
  \caption{The radial distribution of $e_{2}$ component of the eccentricities.
  The disk model is the same as that depicted in Figure \ref{fig:ecc}, and the lines represent the same as those shown in Figures \ref{fig:sigmaave1} and \ref{fig:ecc}.}
  \label{fig:e2}
\end{figure}
To evaluate the eccentricity of the gap's outer edge, we divided the expression as follows:
\begin{eqnarray}
e&=&\sqrt{e_{1}^{2}+e_{2}^{2}},\nonumber\\
e_{1}&=&\left|\frac{r v_{\phi}^{2}}{\mu}-1\right|,\,\, e_{2}=\left|\frac{r v_{r} v_{\phi}}{\mu}\right|,
\label{eq:e2}
\end{eqnarray}
where $e_{1}$ represents the deviation of $v\sub{\phi}$ from the Kepler velocity, and $e_{2}$ corresponds to the contribution from the radial velocity of the gas.
Therefore, it was expected that the $e_{2}$ component would represent the eccentricity of the gap better than $e$ calculated by using Equation (\ref{eq:ecc}).
Figure \ref{fig:e2} shows the radial distribution of the $e_{2}$ component of the disk eccentricity at $t=2000$ orbits.
In this figure, it can be observed that $e_{2}$ components are smaller than $\sim0.02$ in the entire area of the disk for the non-eccentric cases (1 -- $3~\MJ$).
This clearly shows that the double-peak distribution around the planetary orbit is dominated by the $e_{1}$ component of the eccentricity.
On the other hand, $e_{2}$ components were significantly greater outside of the planetary orbit for the eccentric cases, and each case presented with the maximum value of $\sim0.14$.
Therefore, the $e_{2}$ component could be considered a good indicator of the eccentricity of the outer edge of the gap created by the massive planet.

\begin{figure}[]
  \centering
  \includegraphics[]{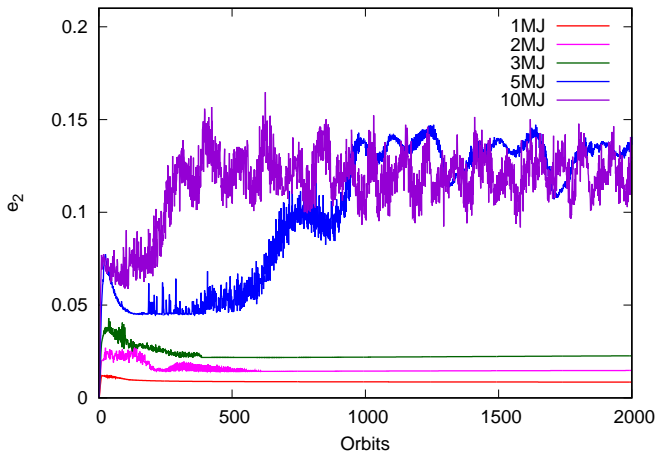}
  \caption{Time evolution of maximum values of $e_{2}$ component of the eccentricities.}
  \label{fig:e2evolve}
\end{figure}
Figure \ref{fig:e2evolve} shows the time evolution of the value at the peak of the radial distribution of the $e_{2}$ component of eccentricity.
The figure suggests that the time evolution can be divided into small $e_{2}$ (with no time variation) and large $e_{2}$ (with time variation).
While considerable variations of $e_{2}$ existed for 5 and $10~\MJ$ cases, the maximum values were similar for the two cases, consistent with the appearance of the eccentric gap structure shown in Figure \ref{fig:sigmamap1}.
It is also shown that $e_{2}$ in the $5~\MJ$ case remained nearly constant during the early phase of evolution when the disk did not reach the quasi-steady-state yet, but it transitioned into time-variable state after $t=500$ orbits.


\subsubsection{Dependence of the gap depth}\label{result_fiducial_depth}


Previous studies conducted on gap formation have shown that the gap depth is determined by the mass of the planet, the aspect ratio, and the viscosity of the disk \citep[e.g.,][]{dm13,fun14,kan15a,kan15b}.
The gap depth can be expressed as
\begin{eqnarray}
\frac{\Sigma\sub{gap}}{\Sigma_{0}}=\frac{1}{1+0.04K},
\label{eq:depth}
\end{eqnarray}
where $\Sigma\sub{gap}$ represents the surface density of the gap region, and the dimensionless parameter $K$ is expressed as
\begin{eqnarray}
K=\left(\frac{\Mp}{\Mstar}\right)^{2}\left(\frac{\hp}{\rp}\right)^{-5}\alpha^{-1}.
\label{eq:K}
\end{eqnarray}
We defined the following three types of gap depth depending on the focus of the study; the averaged surface density within $2~\RH$ from the planet's orbit, the surface density at the deepest point of the gap, and the surface density at $2~\RH$ away from the orbit.
For the gap structure, it is convenient to consider the deepest point of the gap as the gap depth, and it may correspond to the appearance of the gap created by the planet.
In the context of effects of the gap structure on the mass accretion onto the planet, it was important to consider the surface density of the accretion flow toward the planet.
For example, \citet{tw02} conducted local two-dimensional numerical simulations of the flow around the planet embedded in the protoplanetary disk with high resolution.
Their results suggested that flows at $\sim 2~\RH$ away from the planetary orbit mainly accrete onto the planet; therefore, it is useful to define the surface density average within $2~\RH$ from the orbit as the gap depth for the evaluation of mass accretion.

\begin{figure}[]
  \centering
  \includegraphics[]{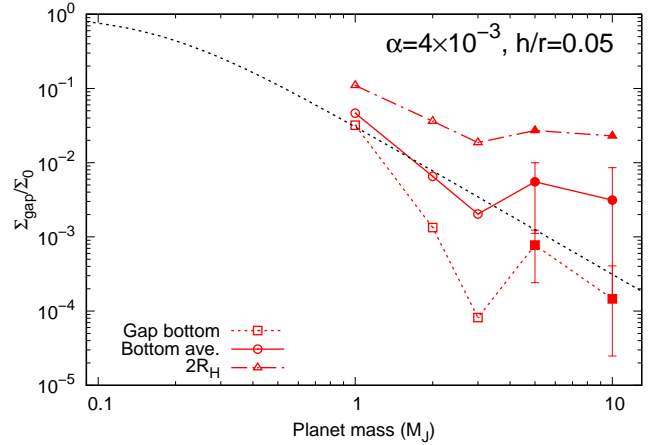}
  \caption{Dependence of gap depth on the mass of the planet.
  The disk parameters are $\alpha=4\times 10^{-3}$ and $h/r=0.05$.
  The solid line shows the averaged surface density within $2~\RH$ from the orbit of the planet, and the dotted shows the surface density at the deepest point of the gap, and error bars represent the time variation.
  The dot-dashed line shows the surface density at $2~\RH$ away from the orbit.
  The open and filled symbols correspond to the non-eccentric and eccentric gaps, respectively.
  The dotted black line corresponds to the empirical relation of the gap depth shown in Equation (\ref{eq:depth}).}
  \label{fig:gapdepth}
\end{figure}
Figure \ref{fig:gapdepth} shows the dependence of the gap depths on the planetary mass.
In the figure, the averaged surface density of the gap within $2~\RH$ from the orbit is shown by using the solid line, and the dotted line shows the surface density at the deepest point of the gap.
Additionally, the surface density at $2~\RH$ away from the orbit is shown by using the dash-dotted line for comparison.
In the cases of 1 -- $3~\MJ$ planets, the averaged gap depth matched well with the empirical relation described by Equation (\ref{eq:depth}).
In cases of 5 and $10~\MJ$ planets, it deviated upward from the relation because the outer edge of the gap became unsteady and the gap was infilled.
It should be noted that the averaged gap depth of the 5 and $10~\MJ$ cases was shallower than that of the $3~\MJ$ case.
Therefore, it was observed that the averaged depth of the gap created by a super-Jupiter mass planet did not deepen as the empirical relation predicted.
The surface density at $2~\RH$ away from the orbit (dot-dashed line) displayed a similar trend; it decreased with the planetary mass in parallel with the empirical relation up to $\sim 3~\MJ$ but bottomed out above $5~\MJ$.

The surface density at the deepest point of the gap (dotted line) exhibited characteristic dependence.
It did not differ from the averaged depth at $\Mp=1~\MJ$, but decreased faster than the averaged depth and reached a minimum at $3~\MJ$.
Above that, the effect of the eccentricity largely increased the surface density at the deepest point.
These results suggested that the gap properties strongly depended on the planetary mass, particularly the eccentric gap created by the heavier planet.

\begin{figure}[htbp]
  \centering
  \includegraphics[]{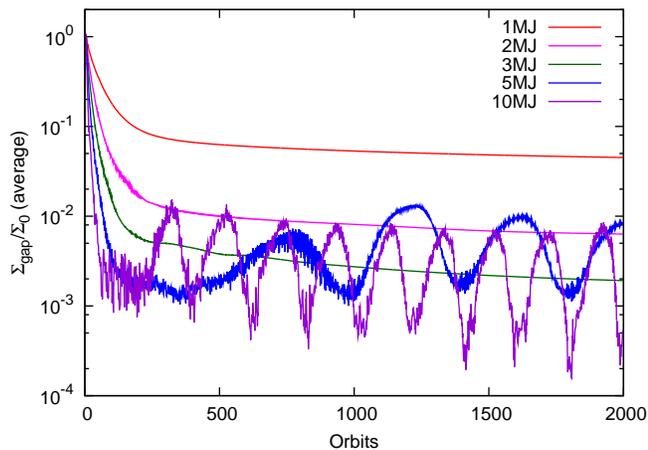}
  \caption{Time evolution of the depth of the gap.
  The surface density of the gap is averaged within $2\RH$ from the orbit of the planet, and is normalized by considering the unperturbed value.}
  \label{fig:gapevolve}
\end{figure}
As shown in Figures \ref{fig:sigmamap_time} and \ref{fig:e2evolve}, the gap properties presented with a substantial time variability when the gap's outer edge gained eccentricity.
Therefore, the gap depths also changed, depending on the relative configuration of the eccentric gap to the planet.
This time variation has been depicted using error bars in Figure \ref{fig:gapdepth}.
Our findings also showed the time evolution of the averaged gap depth in Figure \ref{fig:gapevolve}.
While the averaged gap depths of 1 -- $3~\MJ$ cases did not show time variation, the 5 and $10~\MJ$ cases exhibited substantial variation, and their amplitudes were around an order of magnitude larger.
In the non-steady cases, there was apparent periodicity in the time evolution, and the period was shorter for the heavier planetary masses.
Note that the gap depth shown in the figure was measured at the opposite region of the planet's position to exclude the high-density region around the planet.
Thus, this periodicity corresponded to the relative configuration of the eccentric gap to the planet rather than the variation of the gap depth itself.
In other words, this periodicity was related to the precession of the eccentric gap against the planet; the planet was in a shallower region of the gap when its orbit neared the pericenter of the eccentric gap, and vice versa.
Since the mass accretion rate onto the planet strongly depends on the surface density at the accretion channel around the planet \citep[e.g.,][]{tt16}, the time variation of the gap depth may cause substantial variation in the accretion rate onto a super-Jupiter mass planet.
The possibility of the time-variable accretion of heavier planets due to the disk eccentricity was also described in a study reported by \citet{li21}.
Effects of the eccentricity on the mass accretion rate are discussed in Section \ref{mdot}.


\subsection{Parameter dependence}\label{result_param}


As described by Equations (\ref{eq:depth}) and (\ref{eq:K}), the gap depth also depends on the viscosity and the aspect ratio of the disk \citep[e.g.,][]{dm13,fun14,kan15a,kan15b}.
When the viscosity is high, diffusion of the gas in the vicinity of the gap increases shallowness of the gap, and thus the threshold of the planetary mass for the creation of the eccentric gap may differ.
To clarify this, we performed simulations with different $\alpha$ and aspect ratios.
For the $\alpha$ viscosity, we considered lower viscosity ($\alpha=10^{-3}$) and higher viscosity ($10^{-2}$), and larger aspect ratios ($h/r=0.07$, 0.1) compared to the fiducial cases.


\subsubsection{Transition to eccentric gaps}\label{result_param_transition}


\begin{figure*}[]
  \centering
  \includegraphics[width=5.5cm]{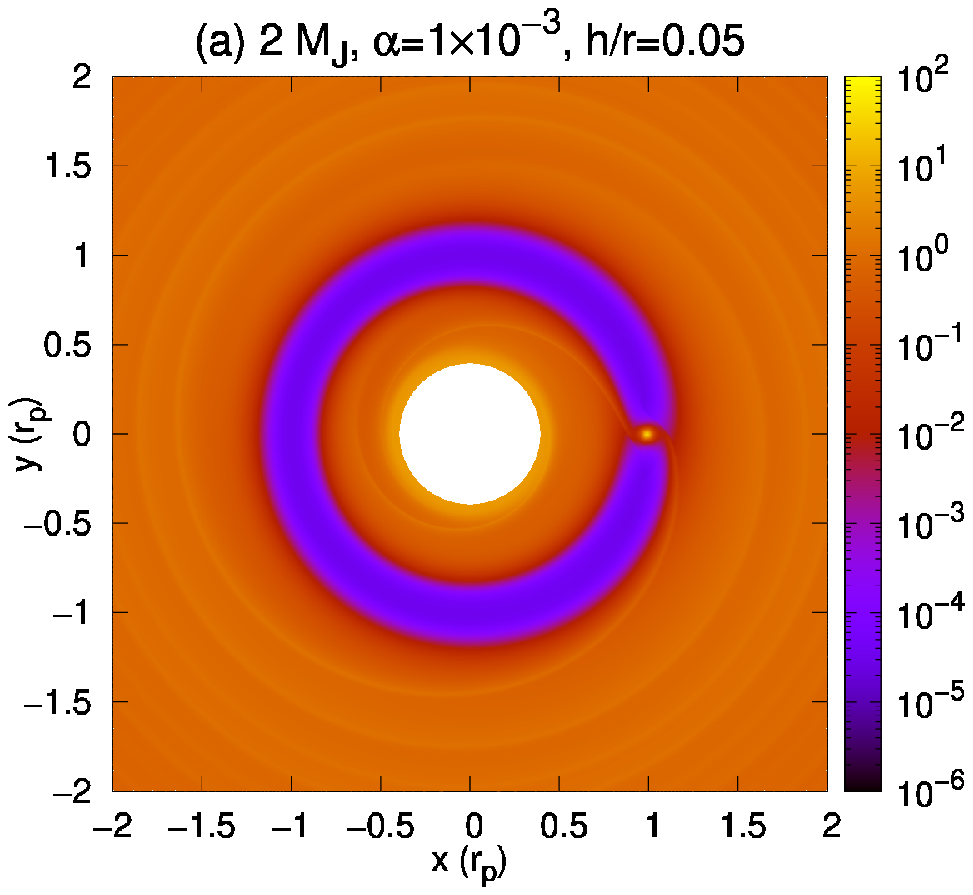}
  \includegraphics[width=5.5cm]{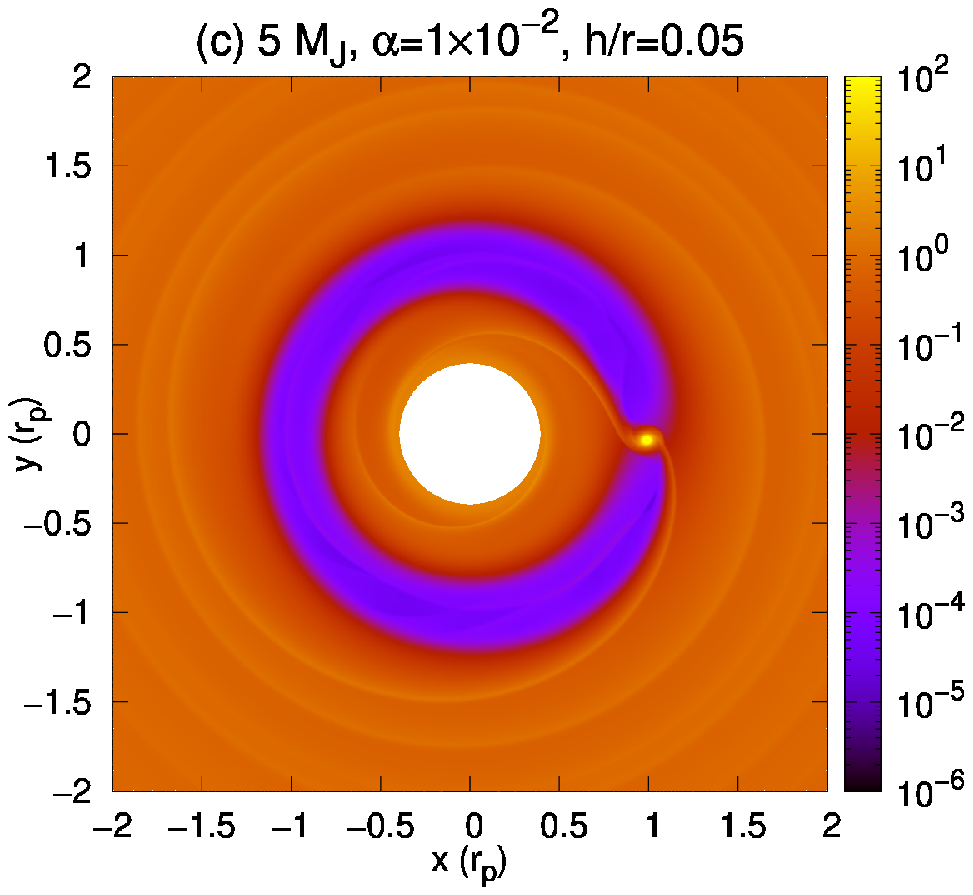}
  \includegraphics[width=5.5cm]{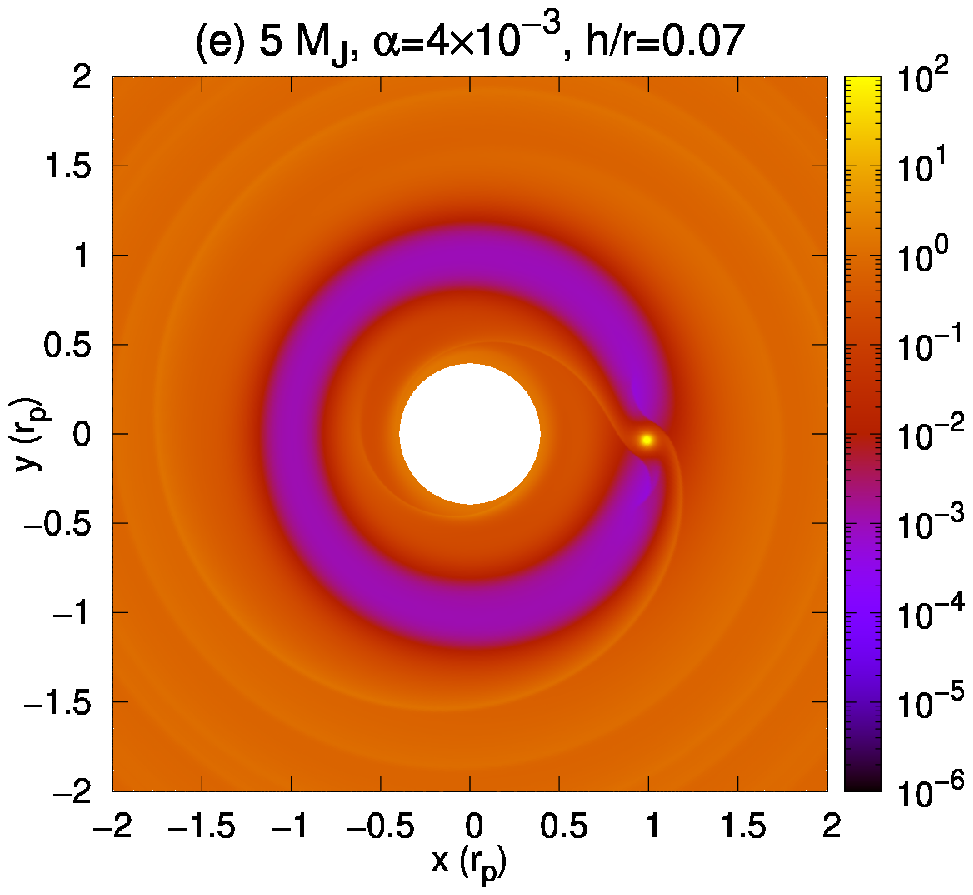}
  \includegraphics[width=5.5cm]{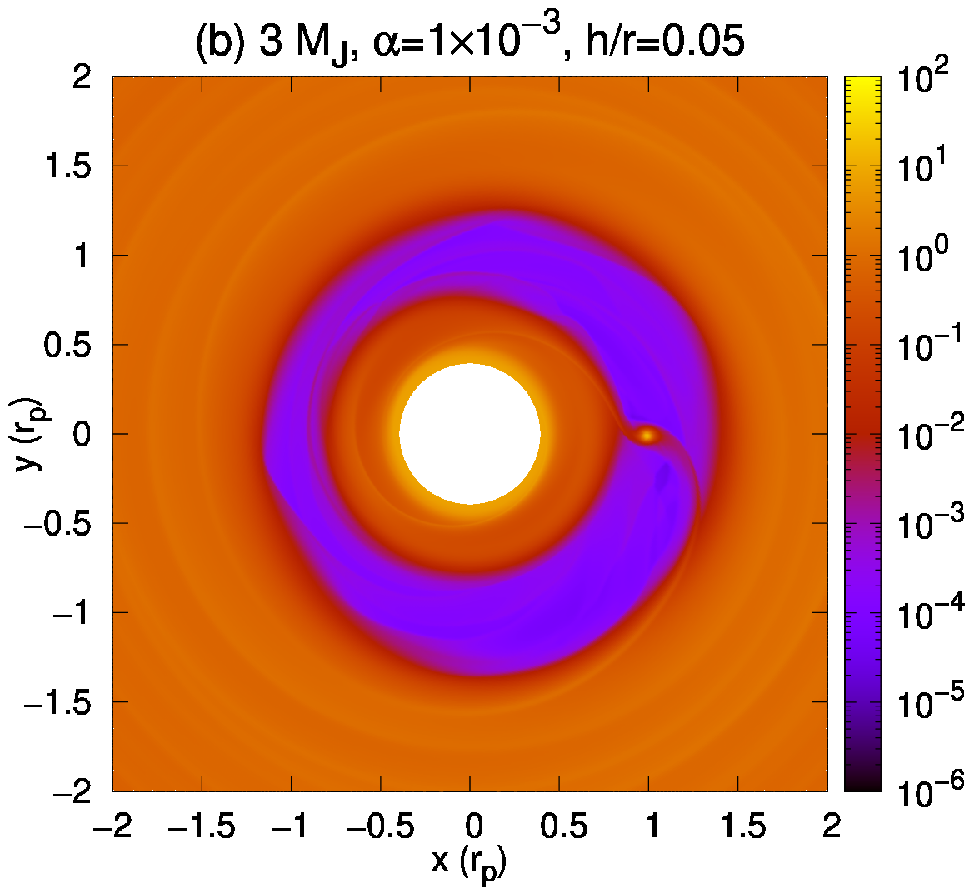}
  \includegraphics[width=5.5cm]{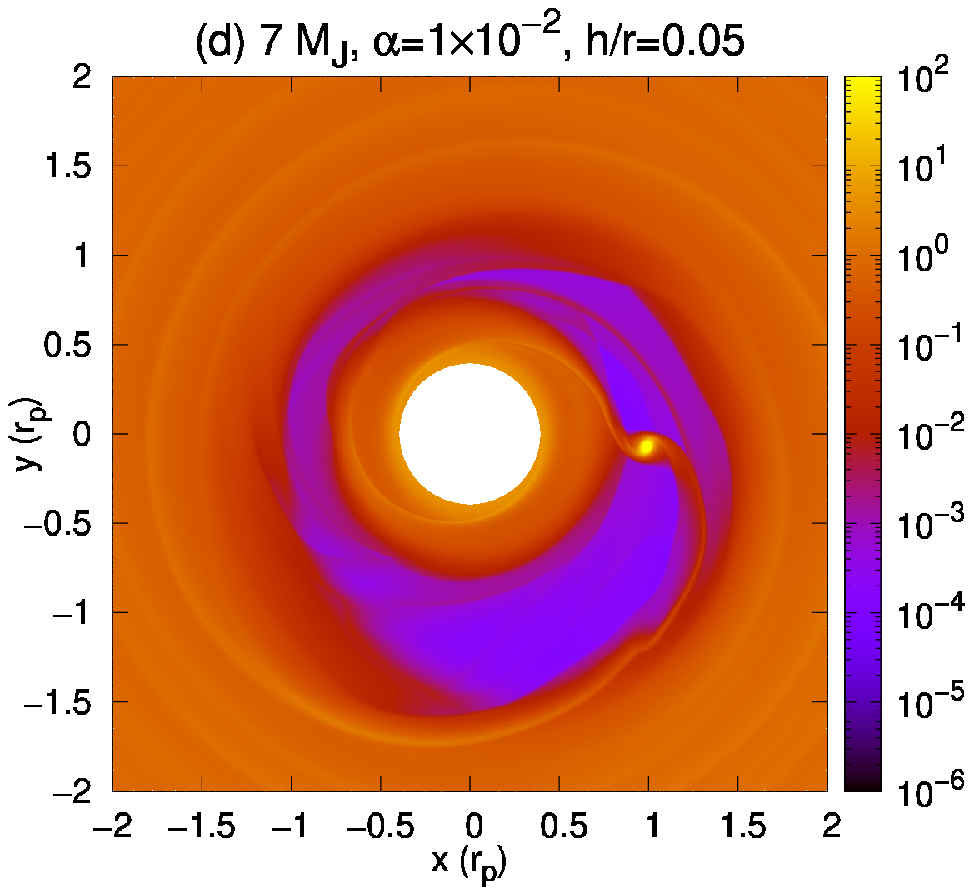}
  \includegraphics[width=5.5cm]{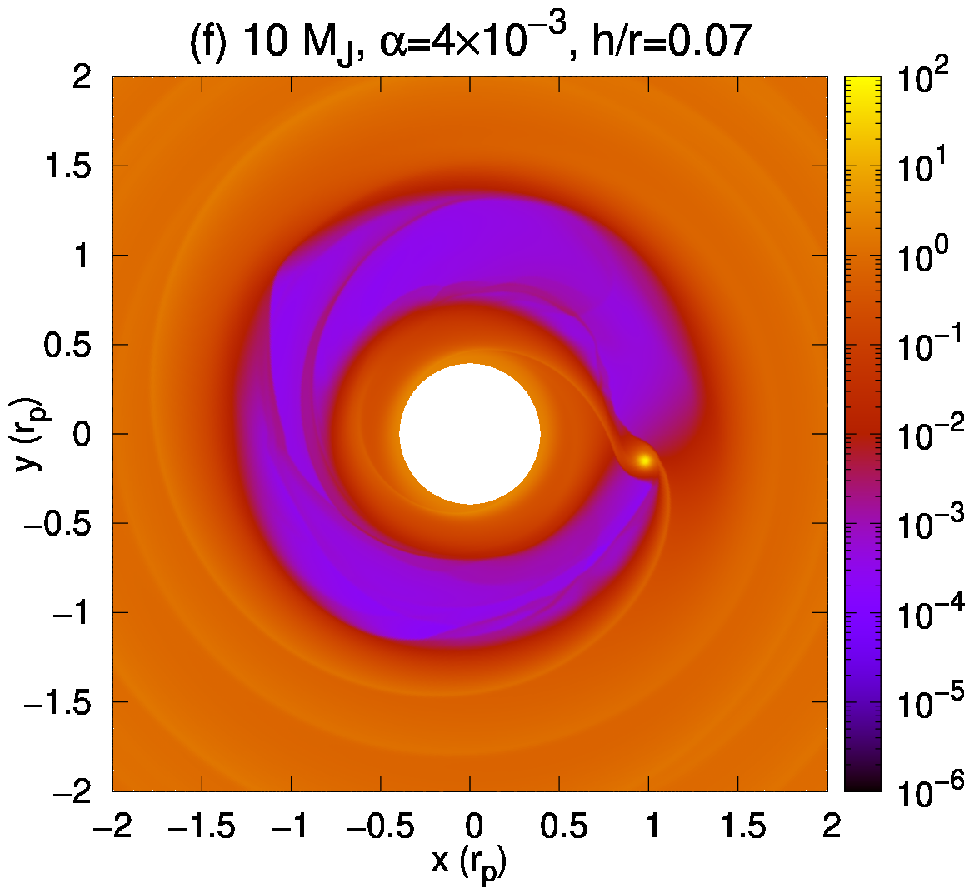}
  \caption{The surface density distribution after 2000 planetary orbits.
  Lower viscosity cases with $\alpha=1\times 10^{-3}$ are shown in (a) and (b), and $\Mp$ values are $2\MJ$ and $3\MJ$, respectively.
  Higher viscosity cases with $\alpha=1\times 10^{-2}$ are shown in (c) and (d), and $\Mp$ values are $5\MJ$ and $7\MJ$, respectively.
  The cases with larger aspect ratio ($h/r=0.07$) are shown in (e) and (f).
  The surface density was normalized by considering the initial (unperturbed) value at $r=\rp$.}
  \label{fig:sigmamap2}
\end{figure*}
Figure \ref{fig:sigmamap2} shows the surface density distributions at $t=2000$ planetary orbits with various parameters.
Panels (a) and (b) represent the lower-viscosity cases of $\alpha=10^{-3}$.
In these cases, the gap created by a $2~\MJ$ planet remained circular, but the gap was deeper compared to the fiducial case shown in Figure \ref{fig:sigmamap1} (b).
The gap created by a $3~\MJ$ planet became unstable, and the outer edge exhibited eccentricity, suggesting that the lower $\alpha$ viscosity enabled a threshold planetary mass to create the eccentric gap to be smaller.
For the higher-viscosity cases (panels (c) and (d)), the gap created by a $5~\MJ$ planet remained circular, and the gap created by a $7~\MJ$ planet displayed significant eccentricity.
The larger aspect ratio of the disk also increaseed the threshold planetary mass; in the cases of $h/r=0.07$ (panels (e) and (f)), the gap remained circular for $\Mp=5~\MJ$, but the outer edge became eccentric for $\Mp=10~\MJ$.
\begin{figure}[]
  \centering
  \includegraphics[]{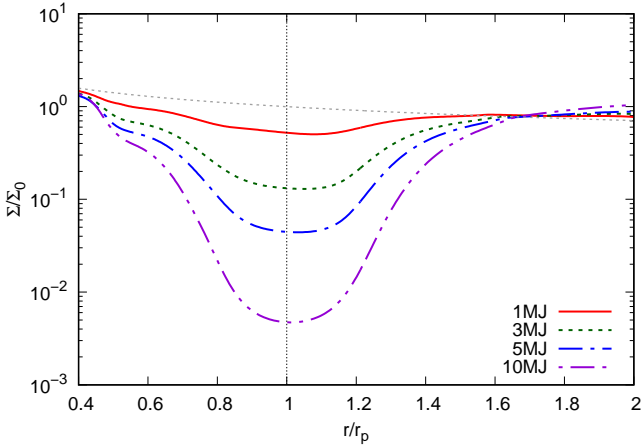}
  \caption{The radial distribution of the surface densities around the gaps for the larger aspect ratio ($h/r=0.1$).
  Lines represent the same as those depicted in Figure \ref{fig:sigmaave1}
  In these cases, $\alpha$ is set to $4\times 10^{-3}$.}
  \label{fig:sigmaave2}
\end{figure}
When the aspect ratio of the disk was much higher, $h/r=0.1$, the outer edge of the gap did not have the eccentricity in any planetary mass we assumed here.
Figure \ref{fig:sigmaave2} shows the radial distribution of the time- and azimuthally averaged surface density of $h/r=0.1$ cases.
In these cases, the gaps were shallower, and the gap profiles were symmetrical with respect to the planet's orbit.
These results suggest that the larger viscosity and the aspect ratio made the gap stable, and the threshold planetary mass to create the eccentric gap increased.

The time evolution of maximum values of the eccentricity of the $e_{2}$ component was same as the fiducial cases; it did not show the time variation and exhibited lower values when the gap was stable, and showed significant time variation and higher values when the gap became eccentric.
We also found that several cases were near a stable and unstable gap border.
For example, in the case of $\alpha=10^{-3}$, $h/r=0.05$, and $\Mp=3~\MJ$, the gap remained almost circular for a while after the start of the calculation, but as the gap deepened, the eccentricity increased and then transitioned into an eccentric state (Figure \ref{fig:sigmamap2} (b)).
This may be caused by slow evolution of the gap depth due to the lower viscosity.
Therefore, the final state of this case was the gap with significant eccentricity.
The case of $\alpha=10^{-2}$, $h/r=0.05$, and $\Mp=5~\MJ$ was also near the boundary, but this case was marginally stable so that the gap outer edge did not have eccentricity (Figure \ref{fig:sigmamap2} (c)).
In this case the surface density inside the deep gap showed small fluctuations not observed in other cases but remained stable at least until $t=2000$ orbits.


\subsubsection{Dependence of the gap depth on the $K$ parameter}\label{result_param_depth}


\begin{figure}[]
  \centering
  \includegraphics[width=7.5cm]{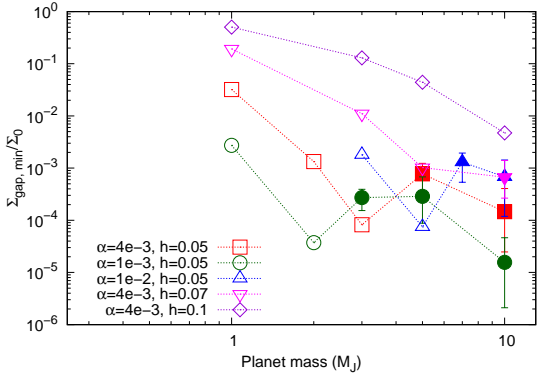}
  \caption{Relation between the mass of the planet and the gap depth.
  The horizontal axis shows the mass of the planet ($\MJ$), and the vertical axis shows the depth of the gap normalized by considering the unperturbed surface density at $r=\rp$.
  The open and filled symbols correspond to the non-eccentric and eccentric gaps, respectively.
  The gap depth shown in the figure was measured at the deepest point, corresponding to the dotted line depicted in Figure \ref{fig:gapdepth}.
  The error bars correspond to the time variation.}
  \label{fig:Msigma}
\end{figure}
The dependence of the gap depth on the mass of the planet in all cases is shown in Figure \ref{fig:Msigma}.
Here, we show the minimum surface density in the gap as a function of the planetary mass.
The results of the fiducial cases (dotted line in Figure \ref{fig:gapdepth}) are also included.
When the gap was steady and not eccentric, the gap depth deepened as the planetary mass increased.
For example, in the cases of $\alpha=4\times 10^{-3}$ and $h/r=0.1$, the surface density of the gap monotonically decreased from $1~\MJ$ to $10~\MJ$.
If the gap was eccentric, the gap depth exhibited increased shallowness and showed time variation as shown by using the error bars in the figure.

\begin{figure}[]
  \centering
  \includegraphics[]{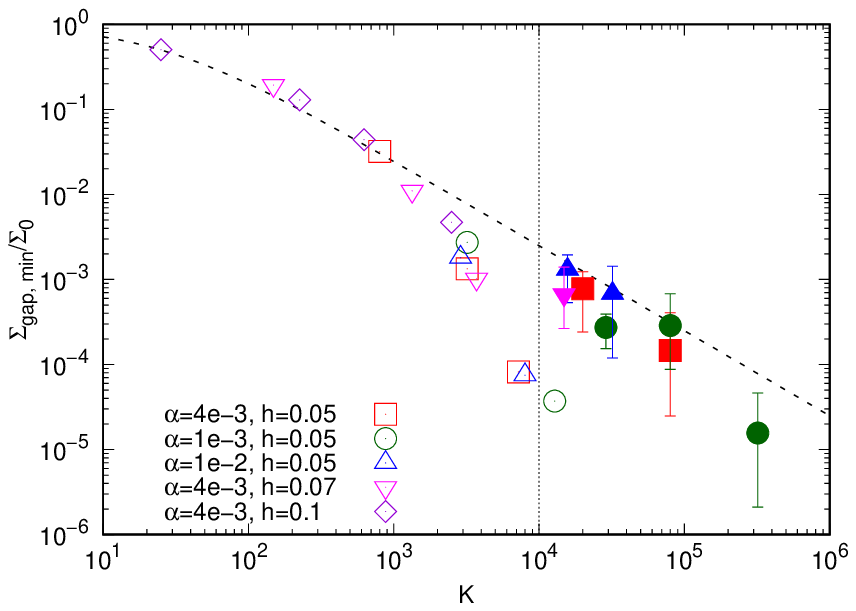}
  \includegraphics[]{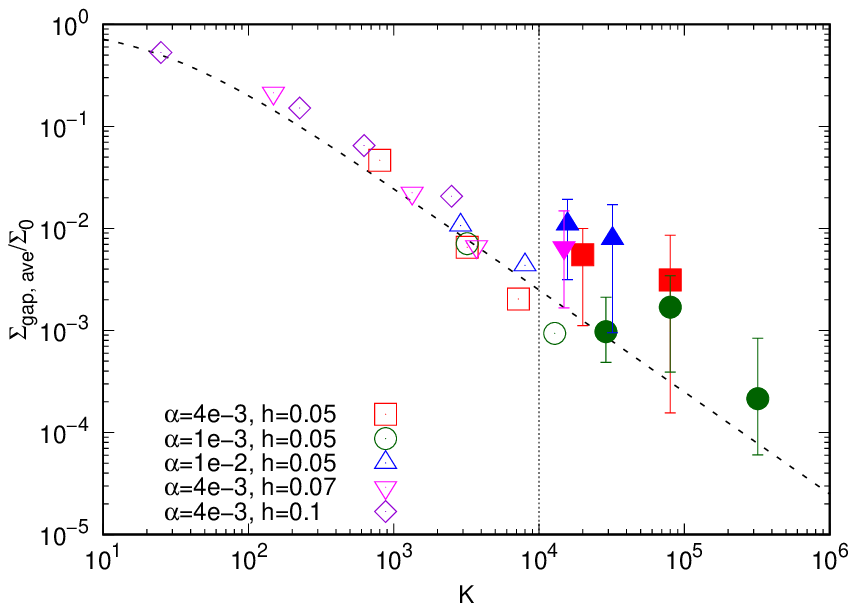}
  \caption{Relation between the parameter $K$ and the gap depth.
  The upper panel shows the surface density at the deepest point of the gap, and the bottom panel shows the surface density averaged within $2\RH$ from the planetary orbit.
  Both values were normalized by considering the unperturbed surface density at $r=\rp$.
  The open and filled symbols correspond to the non-eccentric and eccentric gaps, respectively.
  The black dotted line shows the empirical relation of Equation (\ref{eq:depth}).}
  \label{fig:Ksigma}
\end{figure}
In individual cases, the dependence on gap depth exhibited a marked trend; the surface density inside the gap decreased incrementally until the gap became eccentric and increased when the planetary mass was sufficiently large.
However, overall dependence is unclear in Figure \ref{fig:Msigma} because the gap depth also depends on the viscosity and the aspect ratio, as Equations (\ref{eq:depth}) and (\ref{eq:K}) described.
Since the gap depth can be described by using the dimensionless parameter $K$ \citep[e.g.,][]{kan15a}, it is assumed that the dependence of the gap depth including the gap's eccentricity is also expressed by $K$.
Figure \ref{fig:Ksigma} shows the relation between the gap depths and the parameter $K$.
The upper panel shows the surface density at the deepest point of the gap.
When $K$ was small, the surface density at the deepest point decreased as $K$ increases and presented with a minimum value at $K\sim10^{4}$.
The gap depth around $K\sim10^{4}$ deviates approximately one to two orders of magnitude from the empirical relation.
When $K$ was greater than $\sim10^{4}$ and the gap's eccentricity increased, the gap depth decreased.
This caused a V-shape dependence of the surface density at the deepest point, and then the surface density decreased again in the region of the much larger $K$.
The bottom panel of Figure \ref{fig:Ksigma} shows the relation between averaged gap depth and the parameter $K$.
In this case, the averaged surface density inside the gap decreased as $K$ increased, and the dependence was consistent with the empirical relation up to $K\sim10^{4}$.
Above that value, the average surface density ceased to decrease, displaying the time variation due to the effect of the gap's eccentricity.

These results suggest that development of the gap's eccentricity strongly depends on the properties of the gap structure rather than the gravitational effect of the planet itself.
Additionally, we found that the empirical relation of the gap depth agrees well with the averaged surface density within the gap up to $K\sim10^{4}$, where the gap could maintain the circular structure.
However, the relation underestimated the surface density inside the gap when $K\gtrsim10^{4}$, where an unknown instability occurred and the gap's eccentricity increased.
Therefore, the estimation of the accretion rate onto a super-Jupiter mass planet which strongly depended on the surface density of the gap might be affected.
We also noted that for the lower viscosity scenario, represented by the green filled circle in Figure \ref{fig:Ksigma}, presented with slightly different scaling compared to the other cases.
Therefore, the deviation of the relation might be greater when the viscosity is smaller than the value adopted here.


\subsection{The gap width and the eccentricity}\label{result_width}


\begin{figure}[]
  \centering
  \includegraphics[]{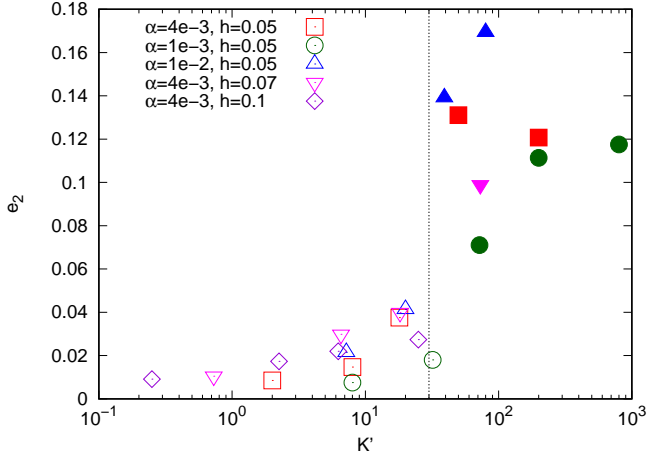}
  \caption{Relation between the parameter $K'$ and the $e_{2}$ component of the eccentricity.
  The $e_{2}$ component of the eccentricity in the figure is indicated by the maximum value at $t=2000$ orbits.
  The vertical dotted line shows $K'=30$.
  The open and filled symbols correspond to the non-eccentric and eccentric gaps, respectively.}
  \label{fig:Kdashe2}
\end{figure}
We have shown that the $e_{2}$ component defined by Equation (\ref{eq:e2}) is a good indicator of the eccentricity of the gap's outer edge (see Figure \ref{fig:e2} and \ref{fig:e2evolve}).
Since the gap depth strongly depended on the planetary mass, viscosity, and aspect ratio via the parameter $K$, it was assumed that the eccentricity of the gap also depended on these parameters.
Here, we used the dimensionless parameter $K'$ which has been defined as \citep{kan17}
\begin{eqnarray}
K'=\left(\frac{\Mp}{\Mstar}\right)^{2}\left(\frac{\hp}{\rp}\right)^{-3}\alpha^{-1},
\label{eq:Kdash}
\end{eqnarray}
to investigate the relationship between the gap's eccentricity and width.
Figure \ref{fig:Kdashe2} shows the relationship between the parameter $K'$ and the maximum value of the $e_{2}$ component of the eccentricity at $t=2000$ orbits.
The figure indicates that the state of the gap's eccentricity can be divided into two types; a gap with low eccentricity (circular) for $K'\lesssim30$ and a gap with high eccentricity (elliptical) for $K'\gtrsim30$.
For $K'\gtrsim30$ scenarios, the gaps exhibit eccentricities of 0.1--0.2.
It was suggested that the case of $\Mp=2~\MJ$, $\alpha=10^{-3}$, and $h/r=0.05$ (the second green filled circle from the left) is almost on the border of the elliptical and circular gap, indicating this might be the marginal parameter for whether the gap's outer edge is stable or not.
Recently, \citet{dem21} have reported that a planet-induced gap will have eccentricity when $K'\gtrsim20$, which is roughly consistent with our results.
They showed that the maximum value of the disk eccentricity transitioned from $K'\lesssim10$ to $K'\gtrsim20$, but our results demonstrated a clearer transition around $K'\simeq30$ because we used the $e_{2}$ component of the eccentricity instead of the eccentricity itself to exclude the effect of the pressure gradient at the gap's edges.

\citet{kan17} showed that the width of the gap could be described using the dimensionless parameter $K'$.
The gap's width can be determined arbitrarily depending on the value of the surface density considered as the threshold of the gap's edge.
However, \citet{kan17} showed that gap widths measured at different thresholds demonstrated similar dependence.
However, the authors used a smaller value for the threshold and elucidated the empirical relation
\begin{eqnarray}
\frac{\Delta\sub{gap}\!\left(\Sigma\sub{th}\right)}{\rp}=\left(0.5\frac{\Sigma\sub{th}}{\Sigma_{0}}+0.16\right)K'^{1/4},
\end{eqnarray}
where $\Delta\sub{gap}$ represents the width of the gap, and $\Sigma\sub{th}$ denotes the threshold value of the surface density for the gap's edge.
Assuming $\Sigma\sub{th}=\left(1/3\right)\Sigma_{0}$ for the threshold, we obtained the relationship between the gap width $\Delta\sub{gap}$ and the parameter $K'$ as
\begin{eqnarray}
\frac{\Delta\sub{gap}}{\rp}\simeq0.33K'^{1/4}.
\label{eq:width}
\end{eqnarray}
As shown in Figure \ref{fig:Kdashe2}, $K'\simeq30$ was thought to be the criterion for the eccentricity.
Thus, Equation (\ref{eq:width}) indicates $\Delta\sub{gap}/\rp\simeq0.76$ for the gap's threshold width.
If the gap is circular and presents a symmetric structure around the planetary orbit, the location of the outer edge of the gap $R\sub{edge}$ can be estimated as $\simeq1.38~\rp$.

\begin{figure}[]
  \centering
  \includegraphics[]{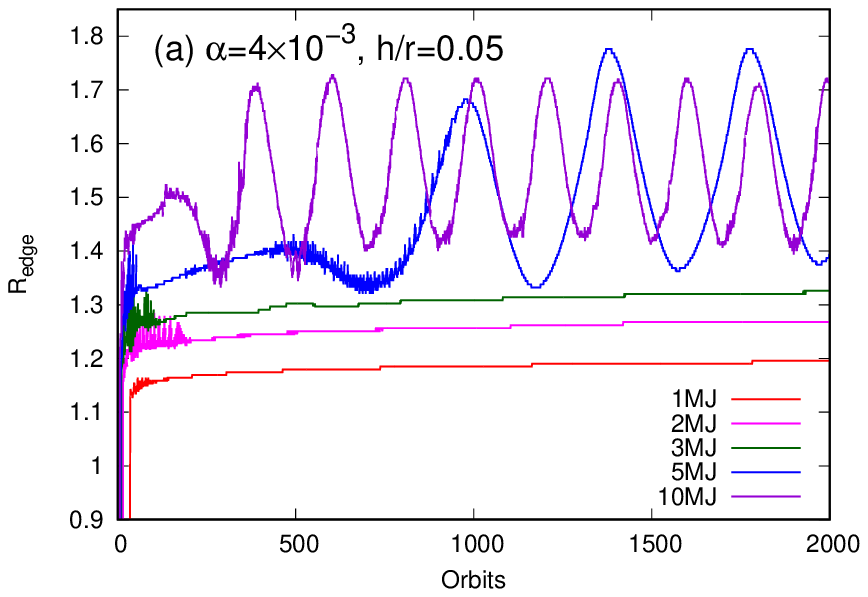}
  \includegraphics[]{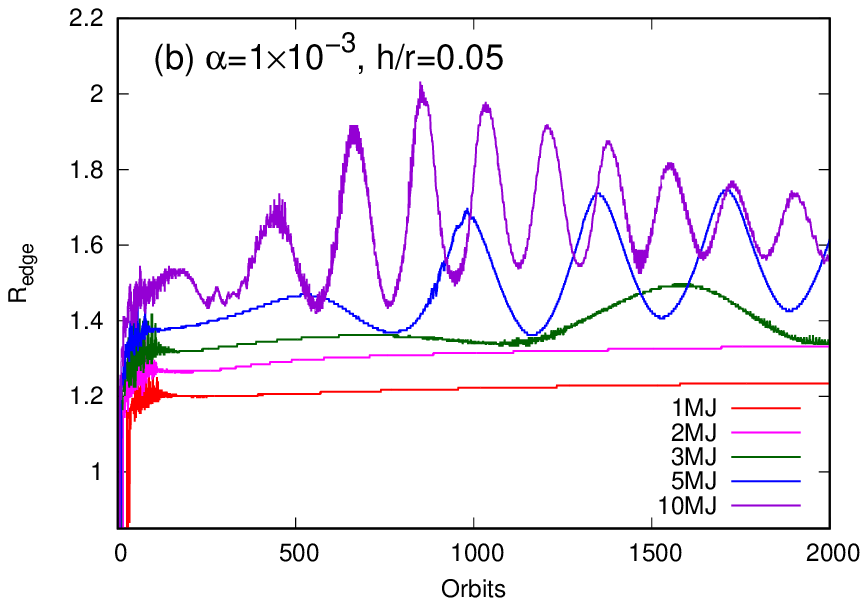}
  \includegraphics[]{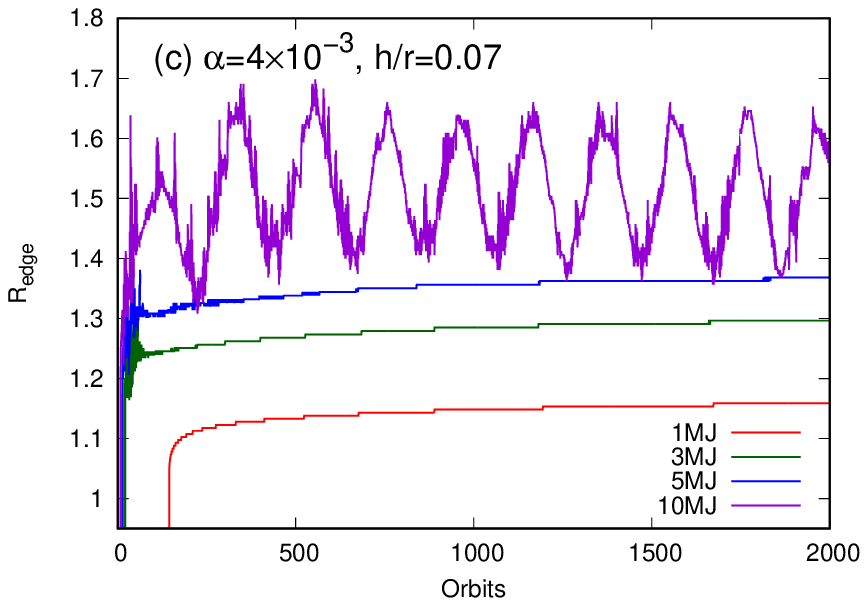}
  \caption{Time evolution of the location of the outer edge of the gap.
  The threshold of the gap edge assumed here is $\left(1/3\right)\Sigma_{0}$.
  The panel (a) is the fiducial cases, (b) is the lower viscosity cases, and (c) is the cases of the higher aspect ratio.
  The vertical axes show the radial position of the gap outer edge, which provides an indication of the width of the gap.}
  \label{fig:gapwidth}
\end{figure}
Figure \ref{fig:gapwidth} shows the time evolution of the location of the gap's outer edge $R\sub{edge}$.
In the cases with circular gaps, $R\sub{edge}$ immediately reached a steady value and it was smaller than $\sim1.38~\rp$, that was derived above.
For scenarios with eccentric gaps, however, $R\sub{edge}$ gradually oscillated after it exceeded $1.38~\rp$, a finding which was consistent with the argument above.
Therefore, these results indicated that gap width might be a critical factor for determining whether the gap's eccentricity increased or not.
When the planet is sufficiently large, and/or the viscosity and the aspect ratio are small, the gap created by the planet widens, leading to an unstable outer edge if the gap's width threshold is exceeded.
We also found several marginal cases in Figure \ref{fig:gapwidth}.
For cases with $\Mp=2~\MJ$, $\alpha=10^{-3}$, and $h/r=0.05$, represented by the magenta line in panel (b), $R\sub{edge}$ was only slightly smaller than $1.38~\rp$.
Therefore, it was suggested that this case barely maintained the circular gap.
On the other hand, for the case with $\Mp=3~\MJ$, $\alpha=10^{-3}$, and $h/r=0.05$ reached the threshold gap width after $t=700$ orbits, and transitioned to an eccentric gap.
Therefore, this case was considered to be barely unstable.

We noted that the time variations of $R\sub{edge}$ shown in Figure \ref{fig:gapwidth} could be considered to represent the relative location of the eccentric gap and the planet, as described in Section \ref{result_fiducial}.
Therefore, minimum and maximum values of the oscillating $R\sub{edge}$ correspond to the pericenter and apocenter of the eccentric gap, respectively.
The eccentricity of the gap's outer edge can be derived from this oscillation.
It is important to compare this method with observations of the gap or cavity in protoplanetary disks that can be curved by super-Jupiter mass planets.
Implications for observations are discussed further in Section \ref{observation}.

\begin{figure}[]
  \centering
  \includegraphics[]{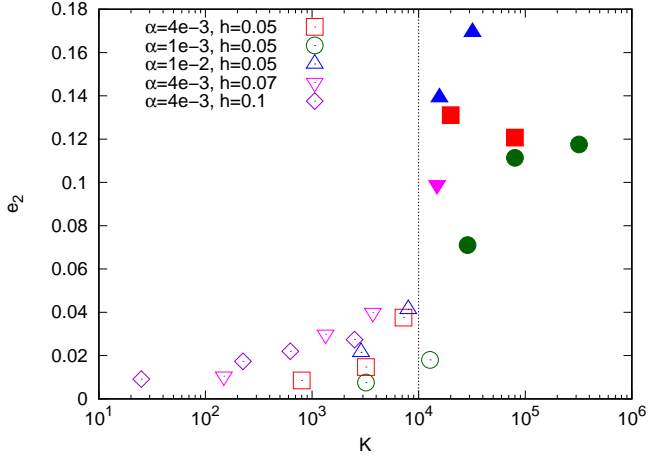}
  \caption{Relation between the parameter $K$ and the $e_{2}$ component of the eccentricity.
  Plots are same as those depicted in Figure \ref{fig:Kdashe2}, and the vertical dotted line shows $K=10^{4}$.}
  \label{fig:Ke2}
\end{figure}
Here, we have investigated the relationship between the $e_{2}$ component and the dimensionless parameter $K'$, an indicator of gap width.
Additionally, we explored the relationship with the $K$ parameter to determine the dependence on the gap depth.
Figure \ref{fig:Ke2} shows the relation between the $e_{2}$ component of the eccentricity and the dimensionless parameter $K$, instead of $K'$.
This relationship clearly transitioned around $K\sim10^{4}$, similar to that shown in Figure \ref{fig:Kdashe2}.
Therefore, it is difficult to determine whether $K$ or $K'$ is a better indicator for the eccentric gaps illustrated in Figures \ref{fig:Kdashe2} and \ref{fig:Ke2}.

We empirically derived a critical planetary mass where the gap induced by the planet has a significant eccentricity.
If we assumed $K'$ as the indicator, the transition from non-eccentric to eccentric gap occurs when
\begin{eqnarray}
q'^{2}\sub{crit}\left(\frac{h}{r}\right)^{-3}\alpha^{-1}=30,
\end{eqnarray}
where $q'\sub{crit}$ represents a critical mass ratio of a central star and a planet.
This can be rewritten as
\begin{eqnarray}
q'\sub{crit}\simeq3.9\times10^{-3}\left(\frac{h/r}{0.05}\right)^{3/2}\left(\frac{\alpha}{4.0\times10^{-3}}\right)^{1/2}.
\end{eqnarray}
Assuming $1~\Msun$ for the central star, a critical planetary mass $M'\sub{p,\,crit}$ is written as
\begin{eqnarray}
M'\sub{p,\,crit}\simeq3.9~\MJ\left(\frac{h/r}{0.05}\right)^{3/2}\left(\frac{\alpha}{4.0\times10^{-3}}\right)^{1/2}.
\label{eq:mcritdash}
\end{eqnarray}
On the other hand, if we assume $K$ as the indicator, the transition occurs when
\begin{eqnarray}
q^{2}\sub{crit}\left(\frac{h}{r}\right)^{-5}\alpha^{-1}=10^{4}
\end{eqnarray}
is satisfied, where $q\sub{crit}$ denotes a critical mass ratio.
Similarly, this equation is rewritten as
\begin{eqnarray}
q\sub{crit}\simeq3.5\times10^{-3}\left(\frac{h/r}{0.05}\right)^{5/2}\left(\frac{\alpha}{4.0\times10^{-3}}\right)^{1/2},
\end{eqnarray}
and a critical planetary mass $M\sub{p,\,crit}$ is written as
\begin{eqnarray}
M\sub{p,\,crit}\simeq3.5~\MJ\left(\frac{h/r}{0.05}\right)^{5/2}\left(\frac{\alpha}{4.0\times10^{-3}}\right)^{1/2}.
\label{eq:mcrit}
\end{eqnarray}
when $\Mstar=1~\Msun$ is assumed.
These two relations shown by Equations (\ref{eq:mcritdash}) and (\ref{eq:mcrit}) are both consistent with the numerical results, and the dependence on the disk parameters is not considerable within reasonable ranges of $\alpha$ and $h/r$.
Therefore, it is difficult to distinguish whether the parameter $K$ or $K'$ is a better indicator for the evolution of gap eccentricity in our calculations.


\section{Discussion}\label{discussion}



\subsection{Surface density at resonances and onset of the eccentricity}\label{resonance}


For the onset of the eccentricity of the gap, it is suggested that interaction established at resonances with the planet in the outer disk plays an important role \citep[e.g.,][]{art92,kd06}.
\citet{lub91} showed that the eccentric instabilities would be driven in disks by resonant interaction with the gravitational potential of the planet.
In this framework, the interaction at the 1:3 resonance with the planet in the outer disk can induce an increase in the eccentricity of the gap when the gap is sufficiently wide \citep{art92}.
If the planet is small, the excitation of the eccentricity by the outer 1:3 resonance will be dampened by other resonances; therefore, the eccentricity of the gap will not increase \citep{kd06}.
It is suggested that the interaction at the outer 1:2 resonance, which is located inside the location of the 1:3 resonance mainly contributes to damp the eccentricity \citep{kd06}.
Therefore, if the gap is wider because of the heavier planet, the smaller surface density at the 1:2 resonance leads to weaker interaction at the 1:2 resonance that will be insufficient to nullify the eccentricity excitation by the 1:3 resonance \citep{kd06}.
To investigate this, we compared the surface densities at the locations of the outer 1:2 and 1:3 resonances, and showed their time evolution.

\begin{figure}[htbp]
  \centering
  \includegraphics[]{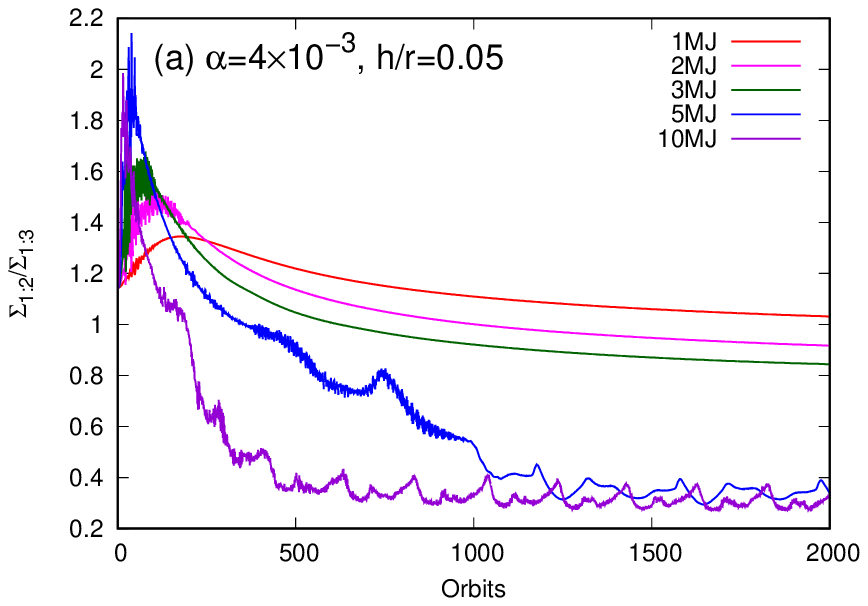}
  \includegraphics[]{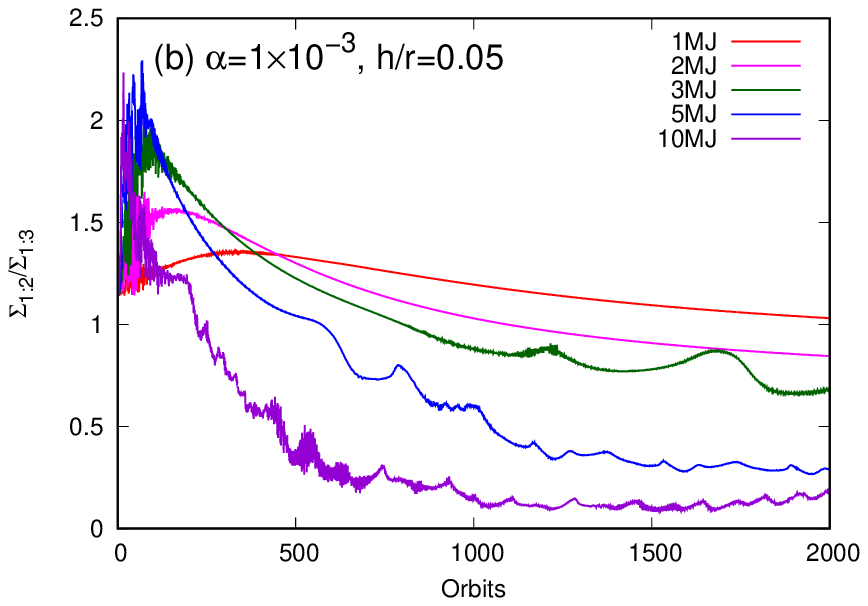}
  \caption{Time evolution of ratios of the surface densities at the resonances.
  The vertical axes show the ratio of the surface densities at the outer 1:2 and 1:3 resonances with the planet.
  Panel (a) denotes the fiducial case and (b) represents the case of lower $\alpha$ viscosity.}
  \label{fig:ratio}
\end{figure}
Figure \ref{fig:ratio} shows the time evolution of the ratio of the surface densities at the resonances.
$\Sigma_{1:2}$ and $\Sigma_{1:3}$ denote surface densities at the location of the 1:2 and 1:3 resonances with the planet.
In all cases, the surface density ratios decreased with the time first as the gap widened; this decrease was mainly attributed to the clearing of the gas at the location of the 1:2 resonance.
In the non-eccentric cases (1 -- $3~\MJ$ in panel (a) and 1, $2~\MJ$ in panel (b)), the decrease of the ratio decelerated and its value remained greater than $\sim0.8$.
In these cases, the disk reached steady-state conditions with the circular gap.
In contrast, the surface density ratio evolved differently in the eccentric cases.
First, the ratio decreased smoothly with time, but it gradually exhibited time variability when $\Sigma_{1:2}/\Sigma_{1:3}$ decreased to less than $\sim0.8$, and the ratio subsequently reduced to a significantly smaller value compared to the non-eccentric cases.
An interesting case is $\Mp=3~\MJ$ in the lower viscosity disk, represented by the green line in panel (b).
In this case the ratio decreased in a manner similar to that observed in non-eccentric gaps until $t\sim1000$ orbits, where it gradually oscillated and transits to the eccentric gap.
This intermediate behavior is also shown in Figure \ref{fig:gapwidth} (b) because the ratio of the surface density at the locations of the resonances strongly reflects the width of the gap shown in Figure \ref{fig:gapwidth}.
This behavior also supports that the condition of the case of $\Mp=3~\MJ$, $\alpha=10^{-3}$, and $h/r=0.05$ is barely unstable.

Figure \ref{fig:ratio} suggests that the ratio of the surface density at the 1:2 and 1:3 resonances is bimodal; $\Sigma_{1:2}/\Sigma_{1:3}$ remains greater than $\sim0.8$ when the outer edge of the gap is stable, and drops to the small value when the gap edge exhibits a significant eccentricity.
This finding is consistent with that reported by \citet{kd06} indicating that the eccentricity of the outer edge of the gap is excited by the tidal interaction at the outer 1:3 resonance and the extent of damping by the interaction at the 1:2 resonance will be reduced when the gap is wide because of the larger mass of the planet.
The relation between the gap width and the eccentricity is also suggested by our finding in Section \ref{result_width}.

However, this bimodality of the ratio may simply be considered to reflect the outcome of instability, and may not be deemed a direct evidence of  instability that causes the eccentricity of the gap.
Previous studies have argued that hydrodynamic instability, such as Rayleigh instability or Rossby wave instability, may occur at the edge of the gap created by the planet \citep[e.g.,][]{li00,ti07,lin12,lin14,kan15b,ono16,kan17}.
For example, it is shown, that the Rossby wave instability can occur at the outer edge of the gap when the viscosity of the disk is low \citep[e.g.,][]{yu10,fu14,zhu14,kan17}.
It is also shown that this instability can be stabilized during disk evolution if the viscosity is considerable \citep[e.g.,][]{lin14,fu14,zs14}.
This is consistent with the finding reported in the present study, indicating that the outer edge of the gap gains eccentricity more easily when the viscosity is low.

\begin{figure}[htbp]
  \centering
  \includegraphics[]{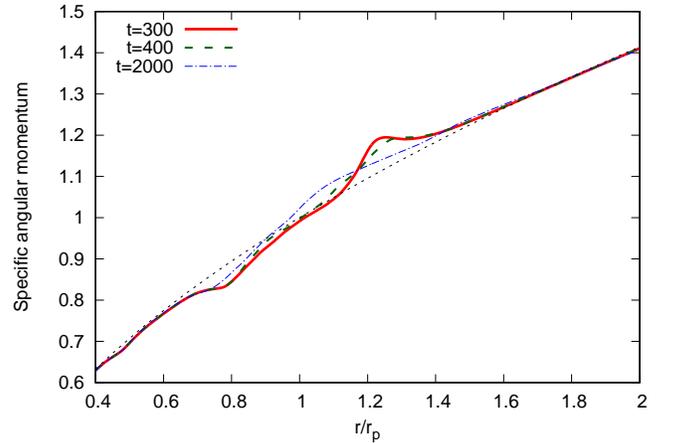}
  \caption{
  The radial distributions of the azimuthally averaged specific angular momentum of the disk around the gap at $t=300$ (solid red), 400 (green-dashed), and 2000 orbits (blue-dotted dash) in the case of $\Mp=5~\MJ$, $\alpha=4.0\times10^{-3}$, and $h/r=0.05$.
  The grey dotted line represents $\propto r^{-1/2}$ as a reference.
  }
  \label{fig:sam}
\end{figure}
For instance, Rayleigh instability occurs when the specific angular momentum $j=r^{2}\Omega$ satisfies $dj/dr<0$ \citep{cha61}.
Figure \ref{fig:sam} shows the radial distributions of the azimuthally averaged specific angular momentum.
In case when $t=2000$ orbits (blue dash-dotted line), the outer edge of the gap already exhibits significant eccentricity and $dj/dr>0$ is satisfied in all regions including the gap.
However, in case when $t=300$ orbits the gap does not exhibit eccentricity, $j$ presents with a local maximum around $r=1.2~\rp$, and $dj/dr$ becomes negative around the outer edge of the gap.
Subsequently, the local maximum of $j$ is not observed as the eccentricity of the outer edge of the gap increases ($t=400$ orbits, shown in green dashed line).
The other eccentric cases followed similar time evolution of the radial distribution of the specific angular momentum.
In contrast, the radial distributions of $j$ in the non-eccentric cases did not show the local maxima around the outer edge of the gap, and $dj/dr$ was positive in the all regions.
This was consistent with the finding reported in a previous work that showed a deep gap induced by a super-Jupiter mass planet could satisfy the condition for the onset of Rayleigh instability \citep{fc16}.
Although our analysis does not suggest that Rayleigh instability is a direct reason of the evolution of the eccentricity, our results imply that a certain type of instability at the outer edge of the deep gap structure may play a key role in the onset of the significant eccentricity.
Therefore, a more detailed analysis is warranted to investigate the origin of the eccentric outer edge of the gap when the mass of the planet is large.


\subsection{Mass accretion rate}\label{mdot}

\begin{figure*}[]
  \centering
  \includegraphics[width=5.5cm]{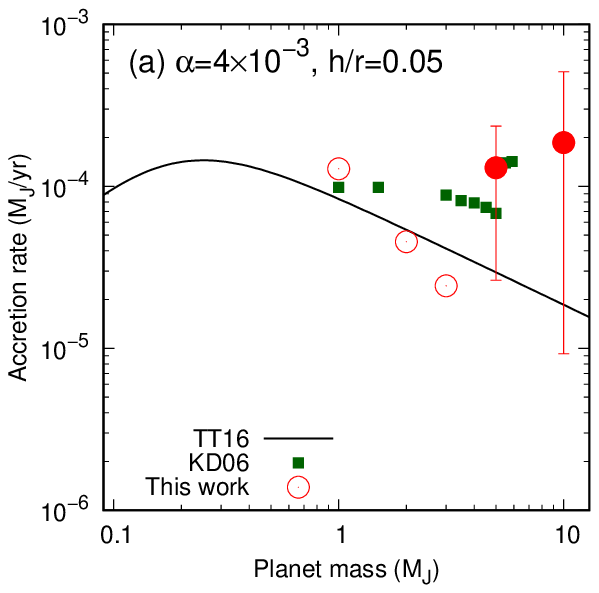}
  \includegraphics[width=5.5cm]{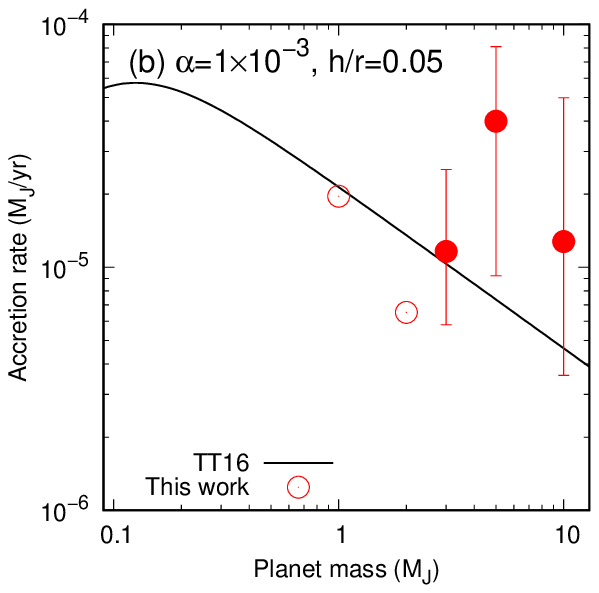}
  \includegraphics[width=5.5cm]{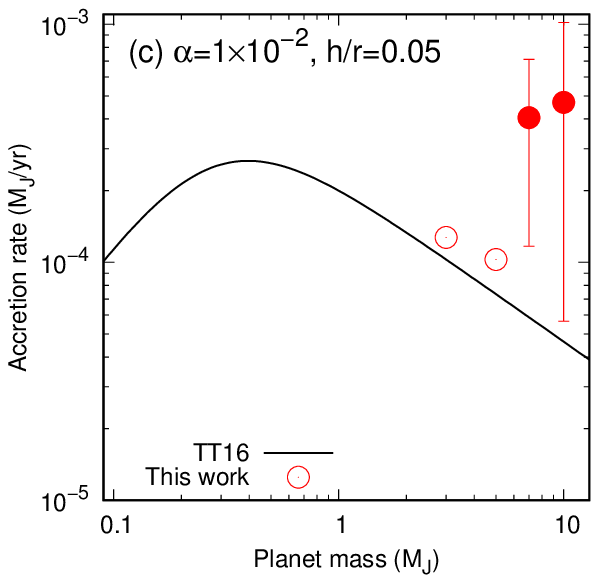}
  \includegraphics[width=5.5cm]{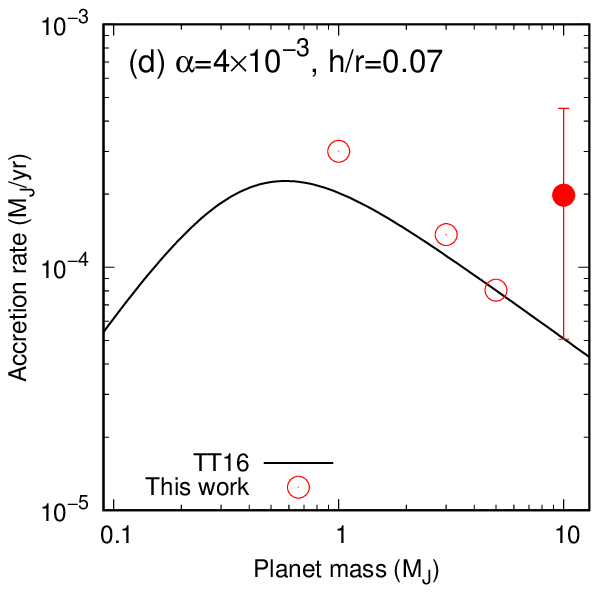}
  \includegraphics[width=5.5cm]{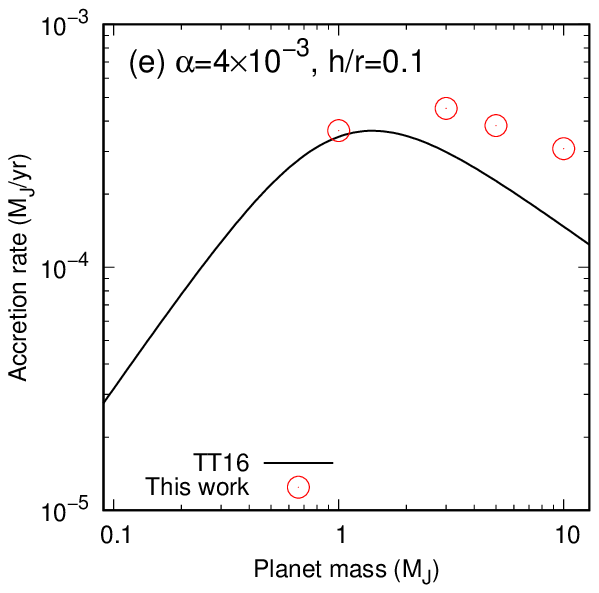}
  \caption{Estimated mass accretion rates onto the planet.
  The horizontal axes indicate the mass of the planet normalized by $\MJ$, and the vertical axes denote the mass accretion onto the planet in units of $\MJ$ per year.
  Here, we adopted a value of the surface density at 5~au in the minimum mass solar nebula \citep{hay85}.
  Our results are shown using the red circles with error bars, and the error bars correspond to the time variation.
  The open and filled symbols correspond to the non-eccentric and eccentric gaps, respectively.
  The green squares in panel (a) denote the results reported by \citet{kd06}, and black solid lines denote an empirical relation of the mass accretion rate as a function of the planet mass reported by \citet{tt16}.}
  \label{fig:mdot}
\end{figure*}
In the simulations conducted in the present study, the mass accretion onto the planet was not included because our main aim was to focus on the detailed structures of the gap and their parameter dependence.
However, since the accretion rate strongly depended on the surface density around the planet, it was possible to estimate the mass accretion rate onto the planet using the outcome of the simulations.
\citet{tt16} introduced a relationship to derive mass accretion rate $\Mdot$ as follows:
\begin{eqnarray}
\Mdot=D\Sigma\sub{acc},
\label{eq:mdot}
\end{eqnarray}
where $D$ denotes an accretion area per unit time, and $\Sigma\sub{acc}$ represents the surface density at the accretion channel.
Dependence of the accretion area $D$ is empirically derived based on local two-dimensional hydrodynamic simulations of the accretion flow around the planet embedded in the protoplanetary disk reported by \citet{tw02}, and it is denoted as
\begin{eqnarray}
D=0.29\left(\frac{\hp}{\rp}\right)^{-2}\left(\frac{\Mp}{\Mstar}\right)^{4/3}\rp^{2}\Omega\sub{p},
\label{eq:D}
\end{eqnarray}
where $\Omega\sub{p}$ represents the Keplerian angular velocity at the orbit of the planet.
Since the surface density at the accretion channel, $\Sigma\sub{acc}$, could be derived from our results of simulations, we obtained the mass accretion rate by combining Equations (\ref{eq:mdot}) and (\ref{eq:D}).
As described in Section \ref{result_fiducial}, the flows at $\sim2~\RH$ away from the planetary orbit mainly accrete onto the planet \citep{tw02}.
We utilized the averaged surface density of the gap within $2~\RH$ from the orbit, adopted as one of the definitions of gap depth, as the surface density at the accretion channel.

Figure \ref{fig:mdot} shows the mass accretion rate onto the planet calculated from surface density of the gap and Equations (\ref{eq:mdot}) and (\ref{eq:D}).
We show an empirical relation of the mass accretion rate onto the planet as a function of the planetary mass for comparison.
This empirical relation is derived from Equations (\ref{eq:mdot}) and (\ref{eq:D}) combined with the empirical relation of the gap depth of Equations (\ref{eq:depth}) and (\ref{eq:K}), as described in the study published by \citet{tt16}.
We have also highlighted the results reported by \citet{kd06} in panel (a) for comparison.

In the fiducial case shown in panel (a), the mass accretion rate decreased with the increase of the planetary mass when the gap was not eccentric ($\Mp\lesssim3~\MJ$).
In this regime, the dependence of the mass accretion rate on the planetary mass was roughly consistent with the empirical relation reported by \citet{tt16}.
When the planetary mass was large and the outer edge of the gap became eccentric, the mass accretion rate increased and deviated substantially from the empirical relation.
As shown in Figure \ref{fig:gapdepth}, the averaged surface density in the gap did not decrease with the planetary mass when the gap exhibited significant eccentricity, because the time-variable flow from the outer disk increased the shallowness of the gap, and considerable amount of the gas could be supplied to the accretion channel.
Since the surface density in the gap does not depend strongly on the planetary mass in this regime, the mass accretion rate will increase proportional to $\Mp^{4/3}$ as described by Equations (\ref{eq:mdot}) and (\ref{eq:D}).
As a result, in our fiducial case, the mass accretion rate onto the $10~\MJ$ planet will be approximately an order of magnitude larger than the value that the empirical relation predicts.
The mass accretion rate also exhibits substantial time variability because the surface density around the planet changes with time due to the eccentricity of the outer edge of the gap, as shown in Figure \ref{fig:gapevolve}.
If the planet is located near the pericenter of the eccentric gap, the surface density at the accretion channel around the planet is relatively large.
Thus, the resultant mass accretion rate is also large.
On the other hand, if the planet is located near the apocenter, the accretion rate drastically decreases because the planet is in lower density region of the gap.
This time variation is shown by using error bars in the Figure \ref{fig:mdot}.
Our results suggest that the amplitude of this time variation of the mass accretion rate can be larger than an order of magnitude.
This substantial variation of the accretion rate may affect observational signal from accreting super-Jupiter mass planets in the protoplanetary disk, for example emission from accreting material around young gaseous planets.

The dependences of the accretion rates in cases of other disk parameters are highlighted in panels (b) to (e).
General behavior is similar to the fiducial case; the mass accretion rate decreases with the planetary mass roughly consistent with the empirical relation during the gap structures remain stable, and shows a considerable increase when the mass of the planet is sufficiently large to induce the eccentricity to the outer edge of the gap.
When the gap demonstrates eccentricity, the mass accretion rate presents with substantial time variation whose amplitude is around an order of magnitude.
The critical planetary mass where the mass accretion rate largely deviates from the empirical relation differs depending on the disk parameter, as shown in Figure \ref{fig:Msigma}, and its dependence can be described by the dimensionless parameter $K$ which is considered the indicator of the gap depth, as shown in Figure \ref{fig:Ksigma}.

\citet{kd06} also highlighted the fact that the eccentricity of the gap would affect the mass accretion rate onto heavier planets, and showed that the dependence of the accretion rate exhibited a marked increase at a critical planetary mass.
The recent numerical study reported by \citet{li21} also showed a similar trend in the mass accretion rate onto the heavier planets.
Our results were consistent with their results as shown in Figure \ref{fig:mdot} (a); the accretion rate decreased with the planetary mass when the gap was stable, and there was an increase in the accretion rate when the mass of the planet was larger than a critical value due to the eccentricity of the gap.
The dependences of the accretion rate differed around $\sim2$ -- $4~\MJ$.
This might be caused by the differences of the numerical setup and the methods adopted to to evaluate the mass accretion rate on to the planet.
Another study conducted on the mass accretion onto heavier planets by \citet{bod13} has shown that the accretion rate continues to decrease rapidly with the increase of the planetary mass, and the dependence does not show any increase such as those reported by \citet{kd06} and our results.
Our calculations support the larger mass accretion rate onto heavier planets, although one must exercise caution while comparing the results of the mass accretion rate because the numerical setup is different in each study; hence, more detailed studies are warranted to evaluate the mass accretion rate onto a super-Jupiter mass planet.

Here, we simply calculated the mass accretion rate based on the empirical relation and our simulation results.
However, an actual accretion rate will be affected by the structure of the flow of the material around the planet \citep[e.g.,][]{tw02}.
When the outer edge of the gap exhibits significant eccentricity, the radial velocity of the gas $v_{r}$ in the gap increases.
This enhancement of the $v_{r}$ in the gap has been depicted indirectly in Figure \ref{fig:e2} via the $e_{2}$ component of the eccentricity denoted in Equation (\ref{eq:e2}).
The relative velocity of the gas to the planet will be larger in this condition.
Thus, this may slightly reduce the accretion rate onto the planet.
Although this effect is thought to be small because the radial velocity of the gas in the gap is smaller or comparable to the Hill velocity $v\sub{H}\sim\RH\Omega$, detailed high-resolution calculations including the effect of the mass accretion onto the planet are necessary to investigate the effect of the gas flow around the planet, especially for cases of super-Jupiter mass planets.

The accretion onto the planet from the protoplanetary disk is important for supply of material to a circumplanetary disk and satellite formation in the disk.
Our results suggest that the supply of the material to a circumplanetary disk around a super-Jupiter mass planet will be greater than the empirical relation predicts, and the rate will exhibit substantial time variation.
Thus, the formation of large satellites around the super-Jupiter mass planets may be affected by the larger mass accretion rate and its time variability.


\subsection{Implications for observations}\label{observation}


Recent development of observation technologies has enabled the observation of young gaseous planets in circumstellar disks that continues to exhibit accretion of material from the disk.
For example, a young T Tauri star PDS 70 is known to host a circumstellar disk \citep{gh02,met04}, and the disk presents a wide gap which is spatially resolved by observations in multiple wavelengths \citep{has12,has15,lon18}.
Two planets, PDS 70b and PDS 70c, have also been discovered in the gap of the disk \citep{kep18,haf19}, and processes of the accretion onto these planets and constraints on the mass accretion rates have been investigated via multiple observations \citep[e.g.,][]{wag18,haf19,chr19,has20,sto20,uya21,zho21}.
Although the estimated masses of these accreting planets present with marked uncertainty that is derived from both observations and consideration of theoretical models used to interpret signals from accreting material, they are thought to be super-Jupiter mass planets \citep[e.g.,][]{kep18,mul18,haf19,wan20}.

\begin{figure}[]
  \centering
  \includegraphics[width=6.0cm]{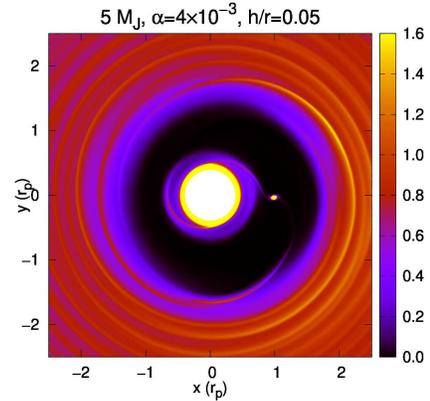}
  \caption{The surface density distribution of the $\Mp=5~\MJ$, $\alpha=4\times10^{-3}$, and $h/r=0.05$ case at $t=2000$ orbits.
  It is identical to Figure \ref{fig:sigmamap1} (d), but the color scale of the surface density is linear, not logarithmic.}
  \label{fig:sigmamaplinear}
\end{figure}
If the gap in a protoplanetary disk that is thought to be induced by an embedded planet exhibits significant eccentricity, it may be used to constrain the mass of the planet and the disk parameters.
Figure \ref{fig:sigmamaplinear} shows an example of the appearance of the eccentric gap induced by a 5-$\MJ$ planet.
$R\sub{edge}$, which denotes the radial distance where the surface density is $\Sigma_{0}/3$, is approximately $1.8~\rp$ at the apocenter of the gap's outer edge and is nearly $1.4~\rp$ at the pericenter.
Therefore, if the semi-major axis of the planet is 5~au, difference between radial distances of the apocenter and pericenter will be $\sim2$~au in this case.
Thus, if an eccentric gap is observed with resolution of sub-au scale, the eccentricity of the outer edge of the gap can be determined and constraint of the planetary mass and/or the disk properties can be calculated.

\begin{figure}[]
  \centering
  \includegraphics[]{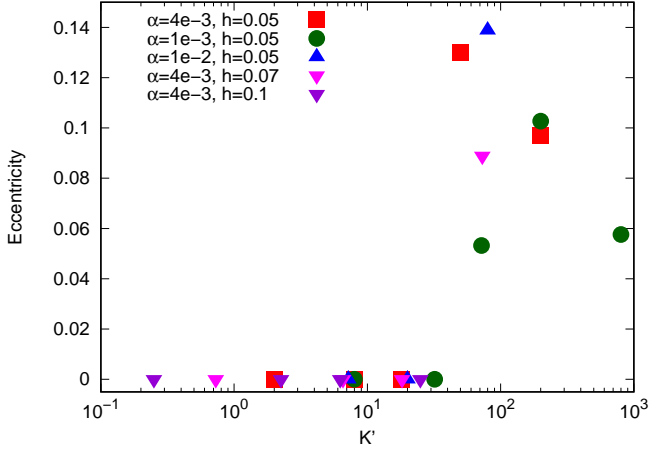}
  \caption{The relation between the dimensionless parameter $K'$ and the eccentricity of the outer edge of the gap calculated by considering the configuration of the gap edge in our simulations.}
  \label{fig:eccfromgap}
\end{figure}
Figure \ref{fig:eccfromgap} depicts the relationship between the eccentricity of the outer edge of the gap and the dimensionless parameter $K'$.
This is similar to the finding illustrated in Figure \ref{fig:Kdashe2}; however, in the figure, the eccentricities are calculated from the values of $R\sub{edge}$ at the apocenter and pericenter of the eccentric gap.
In most eccentric cases, the eccentricities of the outer edge of the gap are $\sim0.1$.
Notably, the eccentricity exhibits substantial temporal variation as shown in Figure \ref{fig:e2evolve}.

In general, observed ring and gap structures in protoplanetary disks are located in relatively outer regions.
The aspect ratio of the gas disk tends to increase with the radial distance; therefore, a heavier planet possesses the ability to induce eccentric gap in the outer region of the disks because the larger aspect ratio strongly suppresses the evolution of the eccentricity of the outer edge of the gap.
For example, if the aspect ratio of the outer region of the disk is 0.1, an object as heavy as a brown dwarf will be necessary to generate the eccentric gap, as shown in Equation (\ref{eq:mcritdash}).
This may be consistent with the fact that observed gaps with significant eccentricities are not frequently encountered.


\subsection{Torque exerted on planets}\label{torque}


\begin{figure}[]
  \centering
  \includegraphics[width=8.5cm]{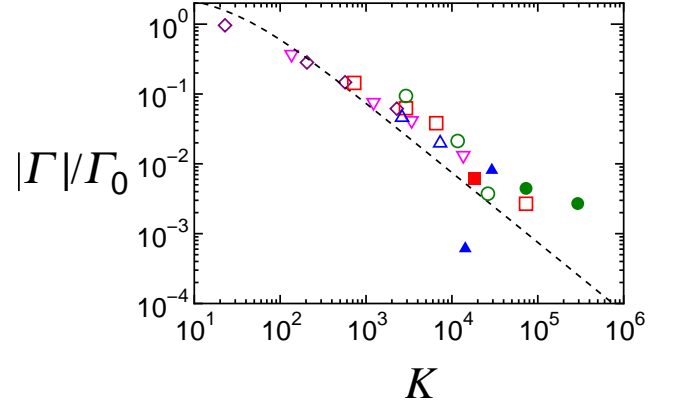}
  \caption{Absolute values of the torque exerted on the planet as a function of the dimensionless parameter $K$.
  Open and filled plots correspond to negative and positive values of the torque, respectively.
  The torque $\Gamma$ is normalized by considering the characteristic torque $\Gamma_{0}$ shown by Equation (\ref{eq:chtorque}), and the black dashed line corresponds to the empirical relation of the torque as a function of the dimensionless parameter $K$, shown by using Equation (\ref{eq:torque}).
  }
  \label{fig:torque}
\end{figure}
Torque exerted on a planet and resultant migration of the planet is beyond the scope of this study; however, we can calculate the torque based on our simulations.
Figure \ref{fig:torque} shows the absolute value of the torque exerted on the planet.
\citet{kan18} derived the expression of the torque exerted on the planet $\Gamma$ as a function of the dimensionless parameter $K$ and expressed it as follows
\begin{eqnarray}
\Gamma=-\frac{1}{1+0.04K}\Gamma_{0}\!\left(\rp\right).
\label{eq:torque}
\end{eqnarray}
$\Gamma_{0}\!\left(\rp\right)$ denotes the characteristic torque exerted on the planet when the gap does not exist, and it is expressed as
\begin{eqnarray}
\Gamma_{0}\!\left(\rp\right)=\left(\frac{\Mp}{\Mstar}\right)^{2}\left(\frac{h}{\rp}\right)^{-2}\Sigma_{0}\rp^{4}\Omega\sub{p}.
\label{eq:chtorque}
\end{eqnarray}
The torque decreases with $K$ and it is consistent with the empirical relation until the achievement of $K\lesssim10^{3}$.
When $10^{3}\lesssim K\lesssim10^{4}$, the empirical relation seems to slightly underestimate the torque.
If $K$ is larger than $\sim10^{4}$, where the outer edge of the gap has the eccentricity, the torque deviates from the empirical relation, and it shows a positive value in certain cases (shown in the filled symbols).
This may imply that a super-Jupiter mass planet existing inside the gap can migrate outward due to the change of the surface density distribution around the orbit.
Although our simulation does not include the orbital migration, the excitation of the eccentric gap is not expected to be affected much by the planetary migration, because the time scale of the growth of the gap's eccentricity is much shorter than the time scale of the orbital migration.
Recent numerical simulations that include the orbital migration have also showed that the eccentric gap is formed for cases of heavier planets \citep{li21,dem21}.

This trend is consistent with recent numerical works that have shown that heavier gaseous planet can migrate outward \citep{duf20,dem21}.
Notably, however, in our numerical setup, the range of the radial computational domain is different from that reported in the previous studies, and thus the results of the torque shown here cannot be compared directly.
Additionally, the torque exerted on the planet can be largely affected by the numerical noise that originates from the region near the inner boundary.
Therefore, the torque and resultant migration of a super-Jupiter mass planet should be considered with caution.


\section{Summary}\label{summary}


We have investigated the detailed properties of gap formation in the protoplanetary disk induced by the super-Jupiter mass planet and its parameter dependence by performing two-dimensional hydrodynamic simulations using FARGO.
We considered the planetary mass, $\alpha$ viscosity, and aspect ratio of the disk as parameters, and their ranges were $\Mp=1$ -- $10~\MJ$, $\alpha=10^{-3}$ -- $10^{-2}$, $h/r=0.05$ -- $0.1$, respectively.
Our results have been summarized as follows:
\begin{enumerate}
\item In fiducial cases ($\alpha=4.0\times10^{-3}$, $h/r=0.05$), the outer edge of the gap induced by the planet remains circular until the planetary mass reaches $3~\MJ$.
Gap eccentricity significantly increases when the mass is heavier than $5~\MJ$ (Figure \ref{fig:sigmamap1}).
\item The planet configuration and gap eccentricity change with time, as shown in Figure \ref{fig:sigmamap_time}.
This suggests that the surface density around the planet varies with time depending on the configuration of the planet in the eccentric gap.
\item The formula for disk eccentricity used in \citet{kd06} (Equation (\ref{eq:ecc})) leads to the exhibition of substantial eccentricities even for circular gaps (Figure \ref{fig:ecc}).
As a better indicator of the disk eccentricity, we introduced $e_{2}$ defined by Equation (\ref{eq:e2}) (Figure \ref{fig:e2}).
\item When the planetary mass is large and the gap is eccentric, the depth is shallower than the empirical relation highlighted by previous studies (Figure \ref{fig:gapdepth}).
Additionally, the surface density inside the gap exhibits substantial time variation in the eccentric cases (Figure \ref{fig:gapevolve})
In case of low-mass planets with circular gaps, on the other hand, the gap depth on the planetary mass is consistent with the previously reported empirical relation.
\item The critical planetary mass for the eccentric gap depends on the viscosity and the disk scale height.
The dependence of the critical mass can be described by considering the dimensionless parameter K, the indicator of the gap depth (Figures \ref{fig:sigmamap2}, \ref{fig:Msigma}, and \ref{fig:Ksigma}).
When $K\lesssim10^{4}$, the outer edge of the gap becomes eccentric, while the gap remains circular for $K\gtrsim10^{4}$.
\item Gap depth and width can be important indicators dividing the eccentric and non-eccentric cases.
The $e_{2}$ component of the eccentricity is strongly correlated with the dimensionless parameter $K'$, which is the important parameter for the gap width (Figure \ref{fig:Kdashe2}).
Since the dependences of the eccentricity on $K'$ and $K$ do not differ significantly, both parameters can be used as the indicator of the onset of the eccentricity (Figure \ref{fig:Ke2}).
\item We examined the surface density evolution at the 1:3 and 1:2 resonances with the planet, which are deemed important for the onset of the disk eccentricity.
We found that when the ratio of the surface densities at the resonances $\Sigma_{1:2}/\Sigma_{1:3}$ becomes smaller than $\sim0.8$, the disk eccentricity starts to evolve (Figure \ref{fig:ratio}).
This result is consistent with the finding reported by \citet{kd06}, however, other mechanisms for growth of the eccentricity such as Rayleigh instability or Rossby wave instability at the outer edge of the gap should be investigated in detail.
\item We evaluated the mass accretion rate onto the planet using the gap depth obtained in our simulations.
We found that the mass accretion rate were larger by approximately an order of magnitude than the empirical relation by \citet{tt16} for cases of eccentric gaps (Figure \ref{fig:mdot}).
This enhancement of the mass accretion rate is consistent with the finding reported by \citet{kd06} and \citet{li21}.
\item Whether the gap exhibits significant eccentricity depends strongly on the disk properties as well as the planetary mass.
Thus, the disk properties of a disk that presents with the eccentric gap can be constrained if the planetary mass is determined by other observations.
\end{enumerate}

We did not include the effect of the mass accretion onto the planet in our simulation because we focused on the detailed structures of the gap induced by a super-Jupiter mass planet and their parameter dependence.
The accretion process and resultant mass growth of the planet will be important for long term evolution during the formation and evolution of the planet; therefore, calculations including the effect of the mass accretion onto the planet will be the next step of our study.
The torque exerted on the planet and its parameter dependence are also important for the final architecture of the planetary systems because they help determining the final orbital distances of the planets.
Our results on the torque are basically consistent with previous results, but further detailed investigation will be necessary because the numerical noise originating from the region near the inner boundary of the computational domain may affect the torque exerted on the planet.
Calculation with wider computational domain is warranted for further studies.

Our simulations presented herein are based on a two-dimensional setup.
This two-dimensional approximation is valid in the range of the planetary masses considered here because the Hill radius of the planet is larger than the scale height of the disk at the location of the planetary orbit.
However, vertical streams will be important for accretion flows from around the gap onto the planet \citep[e.g.,][]{tan12}.
In addition, it is suggested that growth of the gap's eccentricity will be damped when three-dimensional effects are taken into account \citep{li21}.
Therefore, three-dimensional calculations may be warranted if we focus on the detailed accretion process around the planet.


\begin{acknowledgments}
\section*{Acknowledgments}
The authors thank Yi-Xian Chen for fruitful discussions and for sharing their paper in preparation.
We also thank the referee for fruitful comments for improving the manuscript.
The present study was supported by JSPS KAKENHI grant Nos. 18H05438, 17H01103, 19K14779, and 20K04051.
Numerical computations were carried out on Cray XC50 at Center for Computational Astrophysics, National Astronomical Observatory of Japan.
\end{acknowledgments}


\bibliography{ref}{}
\bibliographystyle{aasjournal}



\end{document}